\documentclass[11pt]{emulateapj}

\usepackage{times}
\usepackage[usenames]{color}
\usepackage{amssymb}
\usepackage{graphicx}
\usepackage{epstopdf}
\usepackage{url}

\newcommand{\be}{\begin{equation}}
\newcommand{\ee}{\end{equation}}
\newcommand{\ba}{\begin{eqnarray}}
\newcommand{\ea}{\end{eqnarray}}
\newcommand{\bc}{\begin{center}}
\newcommand{\ec}{\end{center}}
\newcommand{\lsi}{LS~I~$+$61$^{\circ}$303}
\newcommand{\ls}{LS~5039}
\newcommand{\psrb}{PSR~B1259--63}
\newcommand{\hess}{HESS~J0632$+$057}

\newcommand{\chandra} {\emph{Chandra}}

\newcommand{\swift} {\emph{Swift}}
\newcommand{\rxte}  {\emph{RXTE}}

\def \cm2{cm$^{-2}$\,}

\def\ergs {erg\,s$^{-1}$}
\def\ergscm2 {erg\,s$^{-1}$cm$^{-2}$}
\def\ss {s\,s$^{-1}$}

\begin{document}

\shorttitle{\textsc{A magnetar-like event and its nature as a gamma-ray binary}}
\shortauthors{\textsc{Torres et al.}}

\title{\textsc{A magnetar-like event from \lsi \\
and its nature as a gamma-ray binary}}

\author{Diego F. Torres\altaffilmark{1,2}, Nanda Rea\altaffilmark{1}, Paolo Esposito\altaffilmark{3}, Jian Li\altaffilmark{4}, Yupeng Chen\altaffilmark{4}, Shu Zhang\altaffilmark{4}
}

\altaffiltext{1}{Institut de Ci\`encies de l'Espai (IEEC-CSIC),
              Campus UAB,  Torre C5, 2a planta,
              08193 Barcelona, Spain}
\altaffiltext{2}{Instituci\'o Catalana de Recerca i Estudis Avan\c{c}ats (ICREA).}
\altaffiltext{3}{INAF, Osservatorio Astronomico di Cagliari,
localit\`a Poggio dei Pini, strada 54, I-09012 Capoterra, Italy}
\altaffiltext{4}{Key Laboratory of Particle Astrophysics, Institute of High Energy Physics, Chinese Academy of Science, Beijing 100049, China}

\begin{abstract}

We report on the \swift--BAT detection of a short burst from the direction of the TeV binary \lsi, resembling those generally labelled as magnetar--like. We show that it is likely that the short burst was indeed originating from \lsi\ (although we cannot totally exclude the improbable presence of a  far-away  line-of-sight magnetar) and that it is a different phenomena with respect to the previously-observed ks-long flares from this system. 
Accepting as a hypothesis that \lsi\ is the first magnetar detected in a binary system,  we study which are the implications. We find that a magnetar-composed \lsi-system would most likely be (i.e., for usual magnetar parameters and mass-loss rate) subject to a flip-flop behavior,  from a rotational powered regime (in apastron) to a propeller regime (in periastron) along each of the \lsi\, eccentric orbital motion. We prove that  whereas near apastron an inter-wind shock can lead to the normally observed \lsi\ behavior, with TeV emission, the periastron propeller is expected to efficiently accelerate particles only to sub-TeV energies. This flip-flop scenario would explain the system's behavior where a recurrent TeV emission was seen appearing near apastron only, the anti-correlation of GeV and TeV emission, and the long-term TeV variability (which seems correlated to \lsi's super-orbital period), including the appearance of a low TeV-state. Finally, we qualitatively put  the multi-wavelength phenomenology in context of our proposed model, and make some predictions for further testing.

\end{abstract}


\section{Introduction}
\label{OBS}

Besides the highly energetic rotational powered pulsars, there is only a handful of other classes of Galactic objects known to be emitting until GeV -- TeV energies. Some are High Mass X-ray Binaries (HMXB).  

The first identified TeV binary system was a 3.4\,year period binary hosting a 48\,ms radio pulsar in an eccentric orbit around a Be star, namely \psrb\ (Johnston et al. 1992, Aharonian et al. 2005). The emission from this object is thought to be associated with the radio-pulsar wind and its interaction with the radiation field and/or the material surrounding the Be-star. It shows variable radio to TeV emission (Johnston et al. 1999, 2005; Chernyakova et al. 2006, 2009; Tam et al. 2011, Abdo et al. 2011b).
Other two TeV emitting systems, \lsi\, and \ls, are both much closer binaries,  with orbital periods of 26.5 and 3.9\,days, respectively, hosting a very massive star (Be and O types) and a compact object, the nature of which is still unknown. They are also both variable, from the radio to the TeV energy range, often showing their orbital modulation throughout the multi-wavelength energy spectrum (e.g., see, Abdo et al. 2009a,b; Aharonian et al. 2006; Albert et al. 2008, 2009; Torres et al. 2010; Li et al. 2011). Their X-ray emission showed ks-timescale flares (Sidoli et al. 2006; Esposito et al. 2007; Rea et al. 2010; Li et al. 2011), and is, in both objects, characterized by an absorbed power-law spectrum.

Very recently, other HMXBs emitting at high energies have been confirmed. On the one hand, we know of  \hess\ (Aharonian et al. 2007; Hinton et al. 2009). This system again hosts a Be star in orbit with a compact object of unknown nature. This TeV binary has an orbital period of 320\,days (Bongiorno et al. 2011), and again shows radio and X-ray variability (Falcone et al. 2010; Skilton et al. 2009; Rea \& Torres 2011). No GeV emission has been observed from \hess\, yet.  On the other hand, {\it Fermi}-LAT observations of the gamma-ray source 1FGL J1018.6-5856 revealed the presence of periodic modulation with a period of 16.5 days (Corbet et al. 2011). Optical observations found an O6V(f) star, very similar to that of the gamma-ray binary LS 5039, which is coincident with a variable X-ray and radio source, leading to the conclusion that 1FGL J1018.6-5856 is new member of the rare gamma-ray binary class, sharing in this case 
several similarities with \ls. In the latter case, no TeV emission has been reported yet. 
 
But what is the physical nature of these binaries? Two physical scenarios have been put forward (see, e.g., Mirabel 2006). On the one hand, a compact object rotating around the massive companion star drives relativistic jets as result of accretion. The gamma-ray binaries would thus be microquasars (see, e.g, Bosch-Ramon \& Khangulyan 2009 for a review). 
On the other hand,  the compact object could be a rotationally power pulsar which drives a wind (e.g., Maraschi \& Treves 1981). 
There are two flavors of models in this case. The inter-wind models are those where the acceleration of the electron population, which is primary to the gamma-rays observed, is originated in the shock region resulting from the interaction of the pulsar and stellar winds (see, e.g. Dubus 2006). 
Instead, the physical scenario for high-energy photon production in intra-winds models is as follows:  pairs are injected by the pulsar or inner-wind shocks and travel towards the observer, producing Inverse Compton photons via up-scattering thermal photons from the massive star. Gamma-ray photons can initiate an IC cascade due to absorption in the same thermal field in both cases (see, e.g., Sierpowska-Bartosik \& Torres 2008).
Models for gamma-rays with pulsars as compact objects have been recently discussed by Torres (2011).

\begin{figure}
\hspace{-0.4cm}
\includegraphics[width=10.5cm, height=7.8cm]{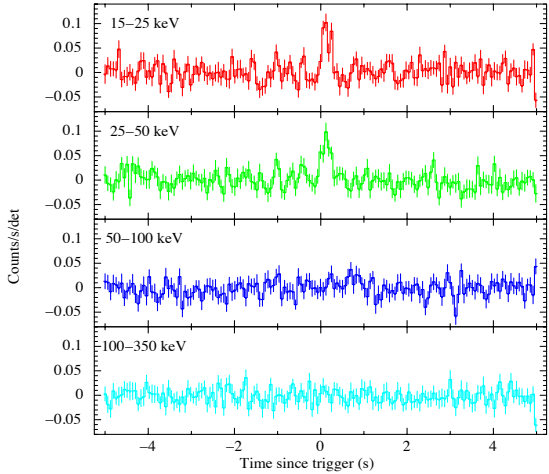}
 \caption{\swift--BAT light curves of the burst detected in the direction of \lsi . 
 }
 \label{batlc}
\end{figure}

\begin{table}
\centering
\caption{Spectral analysis of the \swift--BAT burst.}
\begin{tabular}{@{}lccc}
\hline
 & PL & BB & Brems \\
 \hline
$\Gamma$ & $2.0\pm0.3$ & -- & --\\
$kT$ (keV)& -- & $7.5^{+0.9}_{-0.8} $& $43^{+32}_{-14} $\\
$R$ (Km) & -- & $0.27^{+0.07}_{-0.05} $& --\\
Flux & $5\pm2$&$4.5\pm0.7$ & $4.5\pm0.9$\\
Fluence & $1.4\pm0.6$&$1.4\pm0.2$ &$1.4\pm0.3$ \\
$\chi^2$/dof & 1.29/14 & 1.07/14& 1.22/14\\
\hline
\end{tabular}
\vspace{0.1cm}
\tablecomments{\swift--BAT spectroscopy in the 15--50 keV energy range. Errors are at a 1$\sigma$ confidence level for a single parameter of interest. Fluxes and fluences are in the 15--50 keV energy range and in units of  $10^{-8}$~erg~cm$^{-2}$~s$^{-1}$ and $10^{-8}$erg~cm$^{-2}$, respectively. The blackbody radius is calculated at infinity and for a distance of 2 kpc (which is the distance to \lsi ; Frail \& Hjellming (1991)). \vspace{0.5cm}}
\label{spec}
\label{tabbat}
\end{table}

In the last few decades, two classes of pulsars have attracted an increasing interest: the Soft Gamma Repeaters (SGRs) and the Anomalous X-ray Pulsars (AXPs). They  are two peculiar groups of neutron stars which stand apart from other known classes of X-ray pulsars. In particular, their X-ray luminosities are often larger than the values expected
from tapping their rotational power reservoir ($L\sim10^{35}$\ergs), and they show no evidence for a companion star which could power this strong emission via accretion. Their rotational periods (2--12\,s) and period derivatives ($\sim10^{-13}-10^{-11}$\ss), for most of the $\sim$20 sources known to date, point to a magnetic field of $\sim10^{14}-10^{15}$\,G, which is currently believed to be responsible for their peculiar emission properties (see Rea \& Esposito 2011 and Mereghetti 2008 for recent reviews). At present, in fact, the model which is most successful in explaining SGRs and AXPs emission is the ``magnetar" model: these objects are thought to be strongly-magnetized (isolated) neutron stars emitting across all wavelengths via the decay and the instabilities of their high B-fields (Duncan \& Thompson 1992; Thompson \& Duncan 1993).

The most peculiar and intriguing property of SGRs and AXPs are their outbursts and flares, which are at variance with any other bursting event observed so far in other compact objects. In particular, the unpredictable flaring activity of magnetars can be phenomenologically classified in a few  types:  
\begin{enumerate}
\item  {\em X/$\gamma$-ray short bursts}: these are the most common and less energetic flaring events, they have short duration ($\sim$0.01--0.2\,s), thermal spectra, and peak luminosities of $\sim10^{38}-10^{41}$ erg s$^{-1}$, and they can occur randomly as single events or in a bunch (see recent examples in Rea et al. 2009; Kumar et al. 2010). 
\item{\em Intermediate flares}: they are intermediate both in duration and luminosity between short bursts (1) and the Giant Flares we discuss below. They have durations ranging between $\sim$2 -- 60\,s and luminosity of $\sim10^{41}-10^{43}$ erg s$^{-1}$. Sometimes intermediate flares last longer than the pulsarsÕ spin periods and shows a clear modulation at the star rotational period (see recent examples in Israel et al. 2008; Mereghetti et al. 2009).  
\item {\em Giant flares}: they are by far the most energetic ($\sim10^{43}-10^{47}$ erg s$^{-1}$) Galactic events after supernova explosions. We have detected so far only three of these events, which are characterised by a very luminous hard peak lasting less than a second, and by decaying tails lasting 100--500\,s where the spin period of the neutron star is clearly visible. 
\item {\em Outbursts}: they are enhancements of the multi-band emission of SGRs and AXPs by a factor of 5 -- 1000, with a typical total energy release of $\sim10^{40}-10^{45}$ erg. The increase flux level of the source may last from a few months up to several years (see Israel \& Dall'Osso 2011 and Rea \& Esposito 2011 for a recent review on the flaring activity of magnetars).  
\end{enumerate}
All of the above flaring events are peculiar to, and defining the {\it magnetar} class.

In this paper, we first provide a complete analysis of a magnetar-like short burst detected by the Burst Alert Telescope (BAT; Barthelmy et al. 2005) onboard \swift\, from the direction of the TeV binary \lsi . 
The event prompted several instant notices (Astronomer Telegrams and Gamma-ray bursts Coordinate Network circulars): the SwiftÐBAT trigger was reported by De Pasquale et al. (2008), the \swift--BAT localisation by Barthelmy et al. (2008), and Evans et al. (2008) noticing that a quasi-simultaneous follow up with the \swift--XRT unveiled only one source within the burst error circle: \lsi . Furthermore, brief interpretational comments and reports of multi-wavelength observations already discussing the magnetar-like features of the burst were posted (Dubus \& Giebels 2008, Rea \& Torres 2008, and Munoz-Arjonilla et al. 2008). 

To try understanding the nature of this short burst, the \swift--BAT data were re-analysed, as well as those of various \rxte\, and \chandra\, observations of the field of \lsi . In particular, the data collected by \rxte\, simultaneously with its Proportional Counter Array (PCA) and High-Energy X-ray Timing Experiment (HEXTE; the analysis of these data is presented here for the first time) allows us to explore a  possible alternative origin for the burst, namely, the possibility of it being the result of a spectral evolution in a longer (normal) flare. All of our observational analysis are presented in \S\ref{obs}.
In \S\ref{magnetarbursts} we compare the \swift--BAT burst with those observed from known magnetars, showing in detail the striking similarity with the detection of interest. In \S\ref{MAG} we confront the hypothesis that  \lsi\ might indeed host a magnetar, and start exploring its implications. 
In \S\ref{discussion}, we discuss our results and the proposed model resulting from the previous sections on the light of the multi--band variable emission of \lsi\ reinterpreting all observations from radio to TeV gamma-rays when we place ourselves under the possibility of having at hand evidence for the first magnetar in a binary system.

\begin{figure}
\centerline{\includegraphics[width=6.5cm]{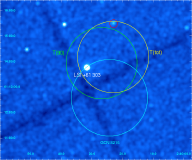}}
\caption{ \swift-BAT positional error circles over-imposed to the {\em Chandra} image of the field (smoothed with a Gaussian function with a FWHM of 3\arcsec ). Cyan, green and yellow circles report on the best position derived by Barthelmy et al. (2008), and our analysis using the T90 and the T(tot) of the burst (see text for details), respectively. The source within the red circle is the {\em Chandra} source discussed in \S\ref{chandra}.}
 \label{batcircles}
\end{figure}


\begin{figure*}
\hbox{
\includegraphics[width=6cm] {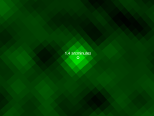}
\includegraphics[width=6cm] {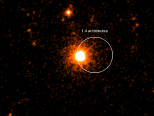}
\includegraphics[width=6cm] {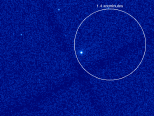}}
\caption{From left to right: {\em Swift}--BAT image of the short burst (see \S\ref{bat}), superimposing the spatial accuracy of our position determination (1.4\arcmin); {\em Swift}--XRT image of all data collected so far on \lsi\, (165\,ks), with the 1.4\arcmin {\em Swift}--BAT error circle superimposed); {\em Chandra} ACIS--I 50 ks image of the field of \lsi\, with the short burst positional accuracy over-imposed.}
\label{images}
\end{figure*}

\section{Observations and Analysis}
\label{obs}

\subsection{\swift -- BAT: a magnetar-like burst}
\label{bat}

The Burst Alert Telescope (BAT; Barthelmy et al. 2005), the hard X-ray detector on board {\em Swift}  (Gehrels et al. 2004), is a highly sensitive coded mask instrument optimized in the 15--150 keV energy range. It was specifically designed to catch and study prompt emission  from gamma-ray bursts and other interesting high-energy transients.

On 2008 September 10 at 12:52:21 UT, the {\em Swift}--BAT triggered on a burst in the direction of the gamma-ray binary \lsi\  (trigger \#324362; De Pasquale et al. 2008; Barthelmy et al. 2008), see Fig. \ref{batlc}.  The data calibration and reduction was performed using the standard BAT analysis software distributed within \textsc{ftools} under the \textsc{heasoft} package (version 6.9), and the latest \textsc{caldb} release (2011--03--03). The mask-tagged (i.e. background-subtracted) counts of the source were extracted from the detector pixels illuminated by the source by using the mask-weighting technique. The mask-weighting factors were calculated by \textsc{batmaskwtevt} using the ground-calculated position of Barthelmy et al. (2008, GCN \#8215; $\rm RA = 40.101^\circ$, $\rm Decl.=61.210^\circ$, about 2.9 arcmin form the BAT on-board position reported by De Pasquale et al. 2008, GCN \#8209; see also Fig.~\ref{batcircles}). 
Fig. \ref{images} present the  {\em Swift}--BAT  image of the burst, together with images of the same region obtained with different instruments, as discussed below.

Mask-tagged light curves were created in the standard 4 energy bands, 15--25, 25--50, 50--100, 100--350 keV (Fig.~\ref{batlc}) at 64\,ms time resolution. The burst is visible in the first two bands (15--25 and 25--50 keV), while no significant excess is observed above 50 keV. 
The total duration of the event in the 15--50 keV band is $T_{\rm tot}\simeq0.31$~s,
while the $T_{90}$ duration is $0.24\pm0.05$~s. 
These values were computed by the \textsc{battblocks} task (based on the Bayesian Block algorithm; Scargle 1998) from a light-curve with 1\,ms bin size.

We extracted a 15--50 keV sky image and performed over the $T_{90}$ duration of the burst a blind source detection with the tool \textsc{batcelldetect}. This script performs a least-square fit of peaks in the map to the BAT point-spread function (a two-dimensional Gaussian) using the local rms noise to weight the pixels in the input map. A single, highly-significant (11.0$\sigma$) point-like source was detected at the best-fit coordinates $\rm RA = 40.1119^\circ$, $\rm Decl.=+61.2322^\circ$ (essentially identical results were obtained using the total duration of the event: $\rm RA = 40.0962^\circ$, $\rm Decl.=+61.2362^\circ$; signal-to-noise ratio 11.0$\sigma$; see Fig.~\ref{batcircles}).

The Point Spread Function fit using {\tt batcelldetect} also yields a formal uncertainty based on the least-square covariance matrix.  At 1$\sigma$ this was 1.06 arcmin for the run using $T_{90}$ (and 1.04 arcmin with $T_{\rm tot}$). However, because neighboring pixels in the coded mask images are inherently correlated, this error is known to be a poor estimator of the true uncertainty for high signal-to-noise detections. Therefore we adopt a more conservative figure of 1.4 arcmin at 1$\sigma$ (90\%: 2.1 arcmin) following the prescriptions of  the BAT calibration reports\footnote{See http://swift.gsfc.nasa.gov/docs/swift/analysis/bat\_digest.html.} and including also a 0.25-arcmin systematic error (see Tueller et al. 2010). This position is consistent with that reported by Barthelmy et al. (2008) (albeit the center of the uncertainty visibly moves) and with that of \lsi\ (the angular separation being $\sim$0.6 arcmin); see Fig.~\ref{batcircles}. No source is detected in the BAT image excluding the burst interval.
The burst spectrum was extracted over the $T_{\rm tot}$ interval (we extracted it also from $T_{90}$ finding consistent results). The results of the spectral analysis are summarized in Table~\ref{tabbat}. To investigate the spectral evolution as a function of time, we computed a hardness ratio, but no spectral variations in the hardness ratio were visible.

\begin{figure}
\centering
\includegraphics[width=9cm,height=6cm] {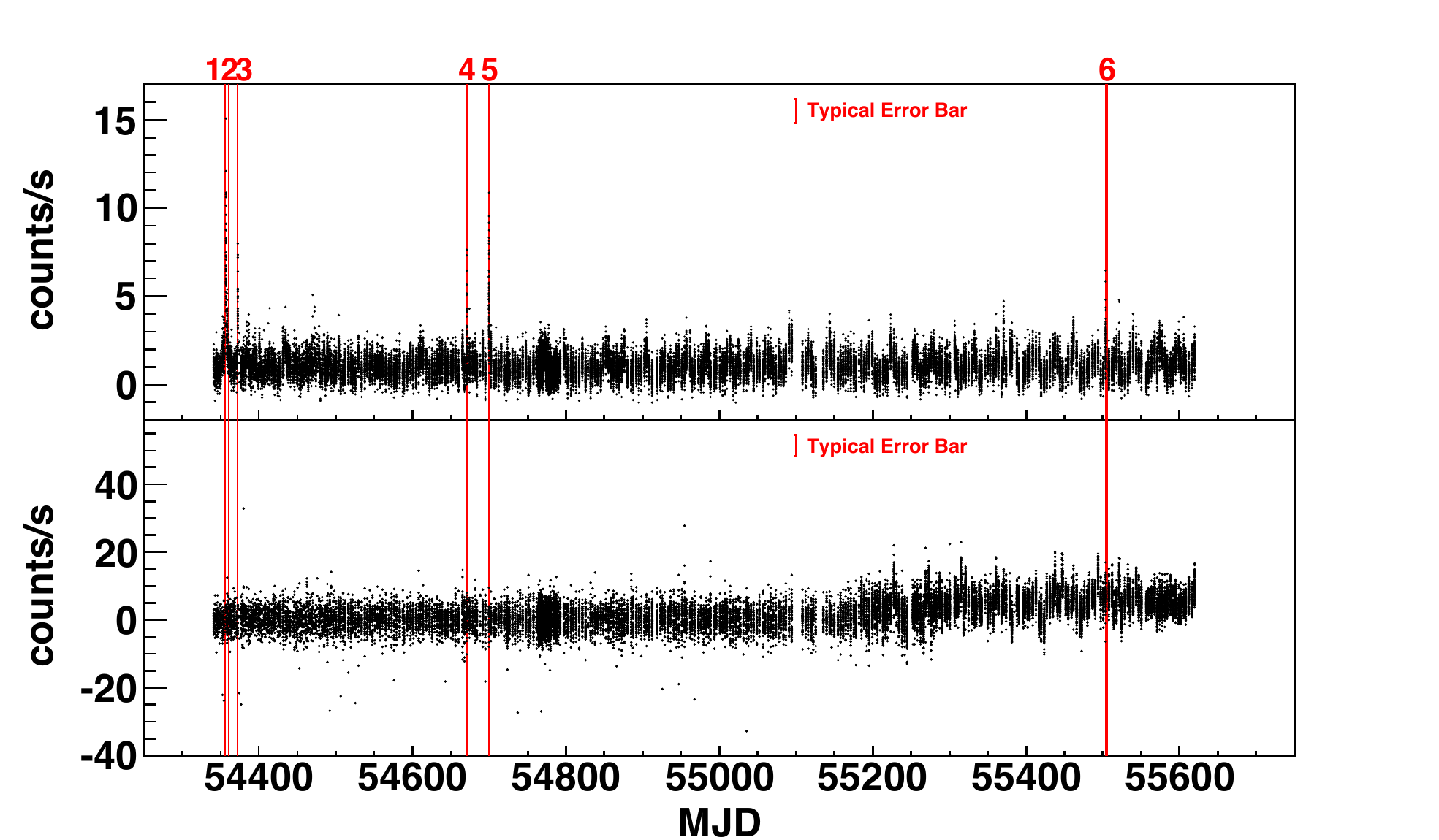}
\caption{{\em Top panel:} \rxte--PCA long-term light-curve of  \lsi\, in 3--10 keV. {\em Bottom panel:} \rxte--HEXTE light-curve in 15--250 keV. The highlighted regions are the 5 flares found in PCA data (Li et al. 2011) plus a sixth additional one we uncovered in this paper as a result of continuous analysis of the new data.}
      \label{flares}
      \label{pcalc}
\end{figure}


\begin{figure*}
\centering
\includegraphics[width=16cm,height=7cm] {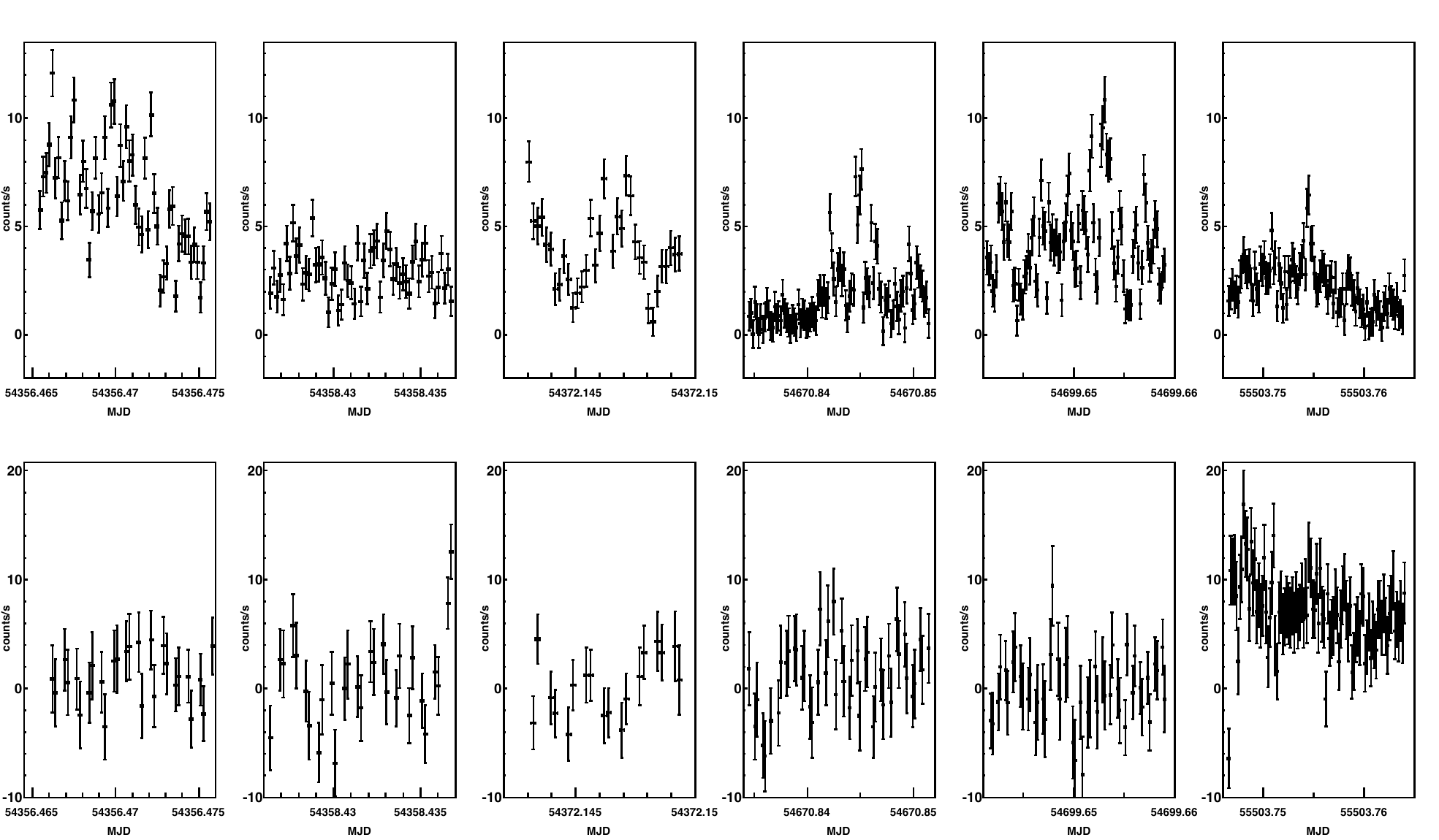}
\caption{Lightcurve of the 6 flares in 3-10 keV (PCA) as observed at 15--250 keV by HEXTE. All lightcurves are binned at 16\,s.}
      \label{flares}
      \label{hextelc}
\end{figure*}

\begin{table} 
\centering
\caption{\rxte--HEXTE counterparts to the soft X-ray flares.}
\begin{tabular}{lcccc}
\hline
Flare & MJD & Average count rate & Significance & reduced $\chi^2$\\
\hline
1  &    54356  &    1.0$\pm$0.5    &     0.3      &          17/26 \\
2  &   54358   &   0.9$\pm$0.5     &    0.2       &         64/28 \\
3  &    54372  &    0    &     0      &          22/17 \\
4  &    54670  &    1.0$\pm$0.4   &     0.3      &          61/45 \\
5  &    54699  &    0   &     0      &          43/47 \\
6  &     55503  &      6.9$\pm$0.3  &      0.5     &            133/94\\
\hline
\end{tabular}
\vspace{0.1cm}
\tablecomments{The average count rate, significance and reduced $\chi^2$ in the 15--250\,keV energy band of the HEXTE data during the ks-timescale flares. }
\label{tabhexte}
\end{table}


\subsection{Analysis of RXTE--HEXTE observations: can the \swift--BAT
 short burst be the result of spectral evolution?}
\label{hexte}

The \rxte--PCA observation nearest to the \swift--BAT burst is located about 6 hr after the burst, with a total exposure time of 1492 s. We have analyzed this observation before, and no unusual behavior (and no flare) was found (see Ray et al. 2008, Torres et al. 2010).

In order to further investigate a possible connection between the short magnetar-like burst seen by \swift--BAT (see \S\ref{bat}), and the typical ks-timescale bursts often observed in \lsi\, (Sidoli et al. 2006; Rea et al. 2010; Smith et al. 2009; Li et al. 2011) we searched for a possible hard X-ray counterpart to the ks-timescale flares observed by the \rxte--PCA, taking advantage of the simultaneous observations performed by the \rxte--HEXTE. The main idea was to search for any short/hard X-ray burst in coincidence with the longer flares observed in the 3--10\,keV energy range (Li et al. 2010), assuming that a strong spectral softening might occur the first seconds after the flare emission. 


\rxte--HEXTE consists of 8 detectors with a total area of about 1600 cm$^{2}$. The 8 detectors are split up in two clusters of 4 detectors each. During normal operations, each cluster is alternately pointed on and off source, generally every 16 or 32\,s, to provide a nearly real-time background measurement. We use here HEXTE data covering the period between 2007 September and 2011 February. The data encompass 418 HEXTE pointed observations, providing a total exposure time of 621 ks on  \lsi.

In the lightcurve analysis of HEXTE we used ``Standard Modes'' data with 16 seconds accumulation time, and built the lightcurve in the 15--250 keV range.
Data reduction was performed using the {\tt HEASoft} tools, and the data were filtered using standard HEXTE criteria. At the end of 2006, the first  cluster (A) of detectors was fixed to
always measure the source of interest and no background measurements were carried out since then. From 2009 December 13, cluster B was fixed to always measure the background and no source detections are available with it.
Additionally, Detector-2 (in the 0--3 numbering scheme) in the second cluster (B) started to function abnormally since 1996 March 6.  Because of this, only Detector 0, 1, 3 in cluster B are used in our analysis till (including) 2009 December 13. After that date, and as suggested by the \rxte\ team, we take data from cluster A as the source and data from cluster B as the background measurements. We select time intervals where the source elevation is $>10^{\circ}$ and the pointing offset is $<0.02^{\circ}$. HEXTE lightcurves were generated, deadtime corrected, and background subtracted using \textsc{rex} script.

In the bottom panel of Fig. \ref{pcalc}, we show the 15--250\,keV \rxte-HEXTE light-curve of  \lsi\ in 16 seconds time bins. For comparison we show in the top panel of the same figure also the 3--10\,keV \rxte--PCA light-curve of  \lsi\ in 16 seconds time bins.
The latter enhances, because of the addition of newer data, the results presented by Li et al. (2011). In particular, it can be seen there that we discovered a sixth flare out of the PCA monitoring.\footnote{
The data set additionally analyzed in this report covers the time span from 2010 September  5 to 2011 February 27 (MJD 55444--55619), and covers nearly another half year of data recently released on the HEASARC website. It includes 50 RXTE pointed observations identified by proposal numbers 95102 and 96102, providing a total exposure time of 70 ks on the source beyond that reported by Li et al. (2011). We follow exactly the same analysis chain as reported by Li et al. (2011) in analyzing the newest data.  The general properties of the new, sixth flare we discover here are very similar to those of the fourth flare reported by Li et al. (2011). Its power spectrum shows no evidence for the existence of any structure.}
   
Since, as we explained above, the overall HEXTE lightcurve is produced with different instruments at different times
--cluster B as source and background before (and including) 2009 December 13 (MJD 55178), and cluster A as source with cluster B as background after that date--
the average of the lightcurve changes, which is obvious to the eye in the bottom panel of Fig. \ref{pcalc}. 
Fitting a constant to the HEXTE lightcurve before MJD 55178 yields an average count rate of 0.11 $\pm$ 0.03, and a reduced $\chi^2 =1.29$ (12915 d.o.f). Data accumulated for more than two years (2007 September -- 2009 December) generates a detection significance for \lsi\ of 4.4$\sigma$ in the 15--250 keV band. The average count rate after MJD 55178  is 4.99 $\pm$ 0.03 and a fit to a constant yields a reduced $\chi^2$ = 1.76 (10745 d.o.f).   
   
To investigate whether there are possible flares in \rxte--HEXTE, we looked at data points with significance above $4\sigma$ over the average count-rate. In the 15--250 keV band, 18 out of 23662 bins points have a significance large than $4\sigma$. None of these corresponds to any of the flares detected by the PCA in the soft X-ray band, and we associate them to statistical fluctuations. A plot of the number of data points with a given count rate would show no deviation from a Gaussian fitting. 
Moreover, we divided the 15--250 keV energy band into 15--60 keV and 60--250 keV and found that all points above $4\sigma$ significance in the larger energy range, 15--250 keV, are no longer that significant in 15--60 keV and/or 60--250 keV. Fig.~\ref{flares} shows, corresponding to the 6 PCA flares marked in Fig. \ref{pcalc}, a zoom of the HEXTE lightcurve. The average count rate, significance, and reduced $\chi^2$ of a constant fit to each flare are listed in Table \ref{tabhexte}, and show that all 6 PCA flares are not significantly detected in the 15--250 keV band. 

If any of the PCA flares would have been preceded by a similar burst to the one detected by \swift--BAT, HEXTE would have seen them.  To show this,  we have simulated a burst observed by HEXTE with  \swift--BAT parameters.
Using the HEXTE response, we found a significance of 4.64$\sigma$ for it.
If the \swift-BAT burst would have been the result of spectral evolution, and 
related to the longer timescales flares usually detected from \lsi, then HEXTE would have spotted flares at least in any of the six instances in which we had PCA and HEXTE simultaneous coverage. This has not happened, HEXTE have not seen any of the PCA ks-timescale flares, and we conclude that their nature is different from the \swift--BAT detection.

\subsection{Chandra ACIS--I: search for a serendipitous magnetar in the field of \lsi}
\label{chandra}

We analysed a $\sim$50\,ks observation of \lsi\, performed with the Advanced CCD Imaging Spectrometer (ACIS) instrument (ObsID 10042) starting on 2006 April 07 22:08:59 (UT),  in VERY FAINT (VF) timed exposure imaging mode (see also Paredes et al. 2007).  The source was positioned on the back-illuminated ACIS-I3 CCD. Standard processing of the data was performed by the Chandra X-ray Center to Level 1 and Level 2 (processing software DS 8.0). The data were reprocessed using the CIAO software (version 4.1.2). 
We used the latest ACIS gain map, and applied the time-dependent gain and charge transfer inefficiency corrections. The data were then filtered for bad event grades and only good time intervals were used. No high background events were detected. The final net exposure time was 49.105\,ks.

We used the CIAO {\tt celldetect} tool to search for sources by summing counts in square cells in the dataset, and comparing the counts to those of "background" cells. At each point where a cell is placed, a signal-to-noise ratio of source counts to background counts is computed. We placed a detection limit of S/N$=$4, and found only one source marginally compatible with the refined position of the \swift--BAT burst (see \S\ref{bat}) at $\rm RA = 40.095081^\circ$, $\rm Decl.=+61.258468^\circ$ (with a 1$\sigma$ error of 1.5\arcsec on the position), with a total of 0.3--10\,kev counts of $50\pm9$ counts (see also Fig.~\ref{batcircles}), translating in a count-rate of $0.0010\pm0.0002$ counts~s$^{-1}$. 
No radio, infrared or optical counterparts have been detected for this source despite the deep archival observations covering this field of view (see Mu\~noz-Arjonilla et al. 2009; this source corresponds to their \#12 of Table 3). Thus, formally we can not exclude that the faint X-ray source detected by {\em Chandra} at the limit of the 1$\sigma$ positional uncertainty of the burst (see Fig. \ref{batcircles}) might be the magnetar responsible for the short burst observed by {\em Swift}-BAT, hence independent from \lsi : Assuming a thermal spectrum of $\sim0.3$\,keV, typical of a magnetar in quiescence (see Rea \& Esposito 2011 for a review), and an absorption column density of $9\times10^{21}$\cm2\, (relative to the whole Galactic value in the direction of the source, from the HI maps from Dickey \& Lockman (1990)) we derived a 0.3--10\,keV observed flux of  $\sim6.1\times10^{-15}$\ergscm2 . Assuming the source is located at the end of the Milky Way, at 10\,kpc distance, the corresponding luminosity would be $\sim2.7\times10^{32}$\ergs , consistent  with it being a magnetar in quiescence. 
Given that the number of TeV binaries is a handful, and the number of magnetars in the Galaxy that have been detected by our experiments is also a handful,
the probability that both are seen at $\sim1.4 \arcmin$ from each other
seems a priori low. A precise number can not be computed without
further assumptions at many levels (population distribution, total number of
sources, etc) which would probably make meaningless the result since
we can never rule out a single random coincidence. The possibility is thus left
open, although it does not seem to be one that would be reasonably
preferred.

\section{The Swift--BAT burst in the context of magnetar emission}
\label{magnetarbursts}

The properties of the burst observed by \swift-BAT\, (a very short duration and a thermal spectrum with a temperature of $\sim7.5$\,keV; see \S\ref{bat}) are typical of magnetars (see Aptekar et al. 2001, Woods \& Thompson 2006; Mereghetti 2008), and at variance with other kinds of frequently observed flares as Type I bursts from X-ray binaries (which last $\sim$100 times longer) or Gamma-ray bursts (which have harder spectra and are much more energetic). In particular, the burst flux (see Table\,\ref{tabbat}) at a distance of 2.3 kpc (as for \lsi) implies a 15--50 keV luminosity of $\sim2\times10^{37}$\ergs .
The luminosity of this burst is in the lower end of the distribution of short bursts from magnetars, in line with the bursts observed in the AXPs (see Gavriil \& Kaspi 2002; Woods et al. 2004) which are usually slightly less powerful (and lasts longer) than the bursts observed from the canonical SGRs (Gogus et al. 1999, 2001; Israel et al. 2008). 


The `relatively' low intensity of the burst, and the single burst that has been found in decades of observations towards \lsi\ and in X-ray all sky surveys, perhaps suggest that the magnetic neutron star producing it has a magnetic field at the lower end of the typical magnetar regimes (hence around $\sim5\times10^{13}$\,G as the case of PSR\,B1846-0258; Gavriil et al. 2008; Kumar \& Safi-Harb 2008), or it is a rather old magnetar (Myrs; see e.g. Perna \& Pons (2011)). 
Note that in the first scenario a high rotational power can be present ($\dot{E}\sim10^{35}-10^{36}$\ergs ), while in the latter scenario the object would be a rather slow pulsar with little energy stored in its rotation (see e.g. Rea et al. 2010 for one of such examples).

However, in both cases  the X-ray emission of \lsi\, can easily be not magnetar-like (e.g. due to resonant cyclotron scattering of a hot surface through a very dense magnetosphere; Thompson, Lyutikov \& Kulkarni 2002). In fact, young magnetars with a relatively low B field and high rotational power have X-ray spectra dominated by non-thermal processes due to particle acceleration and shocks from their strong winds, while old magnetars in quiescence are rather faint ($10^{30}-10^{32}$\ergs ), having dissipated most of their magnetic energy, and emit mostly thermally from their surface.
Thus, from the above considerations and 
what we will explain in detail in the following sections, we believe that in \lsi\, we might be witnessing a high-$\dot{E}$ magnetar with a magnetic field of the order of $10^{13}-10^{14}$\,G.

\section{Magnetar in a binary system}
\label{MAG}

If we entertain the hypothesis that the origin of the magnetar-like event reported here could be \lsi, an effort to understand what are the consequences of the existence of a magnetar in this eccentric binary system is in order. How will the observed multi-wavelength phenomenology be generated in such a case? The rest of this paper is devoted to analyze these questions.

\subsection{Physical radii}

To start considering these issues, we shall introduce several radii (measured from the neutron star),  
which will represent the relative strength of the system's components and compare those with the position of the light cylinder, $R_{lc}$ 
(see, e.g.,  Illarionov \& Sunyaev 1975; Davies \& Pringle 1981, Lipunov et al. 1994).
The latter is 
\be
\label{Rlc}
R_{lc} = \frac{cP}{2\pi} \simeq 4.77 \times 10^9  \left(\frac{P }{ 1 {\rm s} }\right) {\rm cm},
\ee 
where $P$ is the period of the pulsar, and we have adopted for it a scale of 1 s, typical of the observed magnetars.
The Alfv\'en or magnetic radius, $R_m$, 
will be defined as the distance in which the magnetic field starts to dominate the dynamics of the in-falling matter. Thus it can be implicitly 
defined by  the equality between the energy densities, which is attained at $R_m$,
\be
 \frac {B^2}{8 \pi} =  \frac 12   \rho    V_{f}^2.
\label{R_m}
\ee
Here, $V_f$ is the free fall velocity of the accreting matter onto the neutron star of mass $M_{ns}$,
\be
V_ f = \sqrt{   \frac  {2 G M_{ns} } {R} };
\ee
$B$ is the magnetic field in the inner magnetosphere, assumed to be of dipole type with neutron star radius $R_{ns}$
\be
B(R) = B_{ns} \left( \frac {R_{ns} }{ R}  \right)^3;
\label{BB}
\ee
where $R$ is measured from the neutron star center, 
and $\rho$ is the density of the accreting matter. The value of the latter  is defined as 
\be
\rho =  \frac{\dot M _{acc}}{4 \pi R^2  V_f},
\label{rho}
\ee
where $\dot M _{acc}$ denotes the rate at which matter is accreted, say at a distance $r$ from the companion star,
and can in first approximation be obtained using the Bondi-Hoyle-Littleton approach as  (see, e.g., Bondi \& Hoyle 1944)
\be
\dot M _{acc}(r) = \frac14 \dot M _{*} \left( \frac {R_{cap}}{r} \right)^2.
\label{BHL}
\ee
In the latter formula,  $\dot M _{*} $ denotes the stellar mass-loss rate (and again $r$  is measured from the companion, not the neutron star). Finally,
$R_{cap}$ is the gravitational capture radius for a neutron star of mass $M_{ns}$, 
as determined by the relative velocity of the neutron star with respect to the matter, $V_{rel}$, using the orbital 
and the stellar-wind velocity  
\be
R_{cap}  =  \frac{2 G M_{ns} } { V_{rel}^2}.
\label{Rcap-def}
\ee

\subsection{Regimes under the influence of a polar wind}

The stellar wind velocity description will first be assumed having the typical form of a radiatively-driven outflow from a high-mass star 
(e.g., Castor \& Lamers 1979),
\be
V_w = V_0+(V_\infty-V_0)\left(1-\frac{R_*}{r}\right)^\beta \simeq
V_\infty \left(1-\frac{R_*}{r}\right)^\beta,
\label{vel-pol}
\ee
where $R_*$ is the stellar radius,  $V_0 \sim 0.01 V_\infty$, $\beta \sim 1$, and $V_\infty \sim 1000$ km s$^{-1}$ (Lamers \& Casinelli 1999).
Note that in Eq. (\ref{vel-pol}), $r$ is measured from the companion, thus, at $R_m$, we will
have $r(R_m) = d - R_m$ with $d$ being the system separation. In an elliptic orbit of eccentricity $e$, the system separation is given by
\be
d = a (1-e \cos(\epsilon)),
\ee
where $a$ is the semi-major axis and $\epsilon$ is the eccentric anomaly (see, e.g., 
Hilditch 2001). 
%
Considering that 
$d \gg R_m$, we  neglect $R_m$ in favor of $d$ in 
the definition of $r(R_m)$  for the subsequent calculations. 
We have checked, however, that not assuming this approximation would 
correct the results we achieve in  a few percent while significantly complicating the algebra. 
We also consider that the polar wind terminal velocity, $V_\infty$, dominates over the orbital speed (this is explicitly shown below).
Taking this into account, the capture radius is given by  
\ba
R_{cap} & = & \frac{2 G M_{ns} } { V_w^2} \nonumber \\
               &=& 3.73 \times 10^{10}  \left( \frac {M_{ns}} {1.4 M_\odot} \right) \left( \frac{ V_\infty} {10^8 {\rm \; cm \; s^{-1} }} \right)^{-2} \nonumber \\
                && \left( 1- 0.69 \left(\frac{ R_*}{ 10 R_\odot } \right)
         \left(\frac{ a } {10^{12} \; {\rm cm} }\right)^{-1} \right.
 \nonumber \\
 & & \left.         
         \left(1- e \cos(\epsilon) \right)^{-1}  \right)^{-2\beta}
               {\rm cm}.
               \label{Rcappolar}
\ea
%

{We shall consider that a pulsar acts normally, i.e., being rotational powered and driving out a relativistic wind,
if at the capture radius the pressure of the pulsar wind (cosmic rays and magnetic field), given in terms of its spin-down luminosity $L_{sd}$ as 
\be
p_{psr} = \frac{L_{sd} }{ 4 \pi R^2 c} ,
\ee
exceeds the pressure of the stellar wind gas behind the shock ($p_w=\rho_w V_w^2)$, see, e.g.,  Illarionov \& Sunyaev (1975).
In this case, the matter is swept away beyond $R_{cap}$ and the system is so-called ejector (Lipunov et al. 1994).
The pressure condition can be written as 
\ba
L_{sd} & > &  4 \pi R_{cap}^2 c \rho_w V_w^2 .
\ea
This entails a condition onto the pulsar period (for other magnitudes fixed): in order for the pulsar to stop acting as an ejector, the period should be larger than
\ba
\label{out-ejector-condition}
 \!\! \left( \frac{P_{out-ejector}}{ 1\; s} \right) & > & 
3.09 \left ( \frac{ B} {10^{14} \; G } \right)^{1/2} 
 \left ( \frac{\dot M_* } {10^{18} {\rm \; g \; s^{-1}} } \right)^{-1/4} 
    \nonumber  \\
    &&
\times   \left ( \frac{ V_\infty} {10^8 {\rm \; cm \; s^{-1}} } \right)^{3/4} \left ( \frac{a } {10^{12} \; {\rm cm}} \right)^{1/2} 
\nonumber  \\
         & & \times \left( \frac {R_{ns}} {10^6 \; {\rm cm}} \right)^{3/2}   \left( \frac {M_{ns}} {1.4 M_\odot} \right)^{-1/2} 
          \!\! 
          \nonumber  \\
          & & 
          \times \;
          (1 -e \cos(\epsilon))^{1/2} 
\nonumber  \\
         & &     \times  \left( 1- 0.69 \left(\frac{ R_*}{ 10 R_\odot } \right)
         \left(\frac{ a } {10^{12} \; {\rm cm} }\right)^{-1} \right.
            \nonumber  \\
     & &       \left.
    \hspace{1cm}     \left(1- e \cos(\epsilon) \right)^{-1}  \right)^{3\beta/4}  \!\! .
\ea
In these cases, with decreasing radius ($R<R_{cap}$), the pressure of the accreting matter grows as $R^{-5/2}$ whereas the $p_{psr}$ does it as $R^{-2}$. Because of this, matter fall from $R_{cap}$ begins, penetrating within the light cylinder up to the point (if any) in which the quick rise of the dipolar magnetic field in the magnetosphere produces a pressure ($\propto R^{-6}$) able to stop it (i.e., at the Alfv\'en radius $R_m$). The pulsar no longer has a magnetosphere and cannot generate a relativistic wind any longer, leaving the ejector phase.  
In order to move out of this stage, and to ignite the normal (rotationally-powered) pulsar again, an unscathed magnetosphere should be recovered, and thus the condition for it to happen is that the period be smaller than that needed to have 
$R_m= R_{lc}$.  
%
}

The value of $R_m$ can be obtained using previous expressions in Eq. (\ref{R_m}), as
\ba
{R_m} & = & 2.1  \times 10^{10} \left ( \frac{ B} {10^{14} \; G } \right)^{4/7}
\left ( \frac{\dot M_* } {10^{18} {\rm \; g \; s^{-1}} } \right)^{-2/7} 
\nonumber  \\
&& \times \left( \frac{ V_\infty} {10^8 {\rm \; cm \; s^{-1}} } \right)^{8/7} \left ( \frac{a } {10^{12} \; {\rm cm}} \right)^{4/7} 
\nonumber  \\
         & & \times \left( \frac {R_{ns}} {10^6 \; {\rm cm}} \right)^{12/7}   \left( \frac {M_{ns}} {1.4 M_\odot} \right)^{-5/7} (1 -e \cos(\epsilon))^{4/7} 
\nonumber  \\
         & &     \times  \left( 1- 0.69 \left(\frac{ R_*}{ 10 R_\odot } \right)
         \left(\frac{ a } {10^{12} \; {\rm cm} }\right)^{-1} \right.\nonumber  \\
         && \left.
\hspace{2cm} \left(1- e \cos(\epsilon) \right)^{-1}  \right)^{8\beta/7} {\rm cm}.
\label{rmpolar}
\ea

Note that whereas $R_{lc}$ depends (linearly) 
on the neutron star spin period, $R_m$ does not depend on $P$ at all. The line $R_m = R_{lc}$
thus entails a  condition onto $P$. Neutron stars of sufficiently small periods have an unscathed magnetosphere even when in eccentric orbits with high mass-loss rate stellar companions, and thus behave as a normal pulsar. 
In general, in order to re-ignite the pulsar 
the condition over the period (considering all other magnitudes fixed) reads
\ba
 \!\! \left( \frac{P_{into-ejector}}{ 1\; s} \right) &<& 
   4.5 \left ( \frac{ B} {10^{14} \; G } \right)^{4/7} 
 \left ( \frac{\dot M_* } {10^{18} {\rm \; g \; s^{-1}} } \right)^{-2/7} 
    \nonumber  \\
    &&
\times   \left ( \frac{ V_\infty} {10^8 {\rm \; cm \; s^{-1}} } \right)^{8/7} \left ( \frac{a } {10^{12} \; {\rm cm}} \right)^{4/7} 
\nonumber  \\
         & & \times \left( \frac {R_{ns}} {10^6 \; {\rm cm}} \right)^{12/7}   \left( \frac {M_{ns}} {1.4 M_\odot} \right)^{-5/7} 
          \!\! 
          \nonumber  \\
         & &     \times \;
          (1 -e \cos(\epsilon))^{4/7} 
\nonumber  \\
         & &     \times  \left( 1- 0.69 \left(\frac{ R_*}{ 10 R_\odot } \right)
         \left(\frac{ a } {10^{12} \; {\rm cm} }\right)^{-1} \right.
            \nonumber  \\
     & &       \left.
    \hspace{1cm}     \left(1- e \cos(\epsilon) \right)^{-1}  \right)^{8\beta/7}  \!\! .
\label{Rm=Rlc}
\ea
{Using the fiducial values in Eqs.  (\ref{out-ejector-condition}) and (\ref{Rm=Rlc}) }
we see that 
a neutron star with an spin period usually measured for magnetars
would be right at the transition range. Small changes in the mass-loss rate, for instance, can make 
the neutron star  flip-flop from accreting within the magnetosphere to behaving as a 
rotational powered pulsar.

\begin{figure}[t]
\hspace{-.90cm}
\includegraphics[angle=-90,scale=0.38]{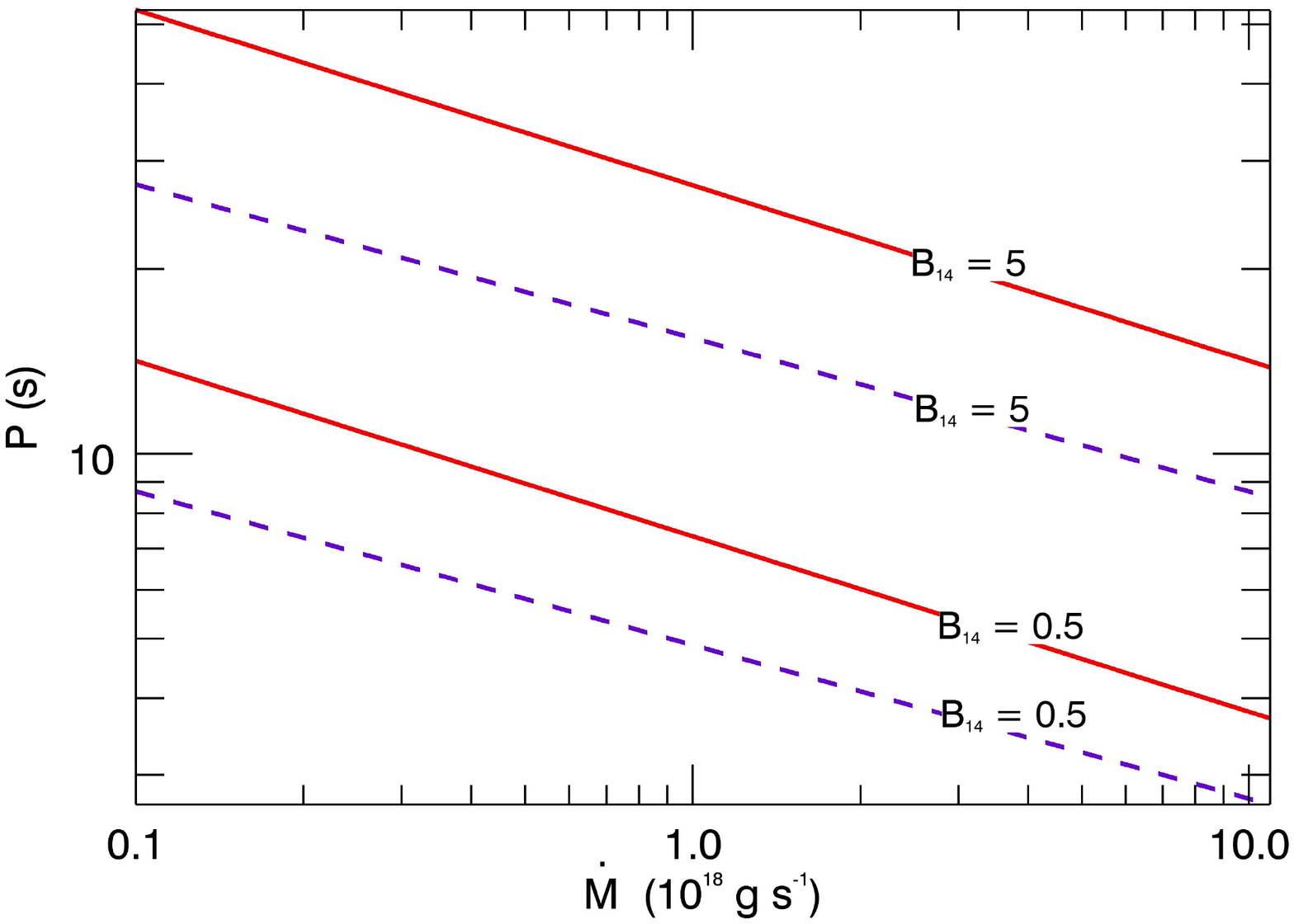}
\vspace{-0.5cm}
\caption{Comparison of the out of ejector conditions written in Eq. (\ref{out-ejector-condition}) --in violet--
and the constraint of the equality $R_m = R_{lc}$, given by Eq. 
(\ref{Rm=Rlc}) --in red. The latter is always a more restrictive condition onto $P$, for equal values of the other
magnitudes involved. Two cases are shown for 
surface magnetic fields of 5 $ \times 10^{13}$ G and 5 $ \times 10^{14}$ G. The eccentricity is assumed equal to zero and
the \lsi\ semi-major axis is adopted in this example. }
\label{compe}
\end{figure}

To simplify the reasoning that follows, we will consider that the condition 
$R_m = R_{lc}$ not only establishes the {\it into-ejector}  condition over the period $P$, but also the
{\it out-of-ejector} regime. Indeed, the out-of-ejector condition obtained with Eq. (\ref{out-ejector-condition}) 
is easier to fulfill (i.e., slightly smaller periods can fulfill it, for equal values of the other
magnitudes involved) than that obtained with Eq. (\ref{Rm=Rlc}), whereas the shape of the constraint is similar.
Fig. \ref{compe} compares these two constraints over $P$, differing by a factor of $\sim 1.57$ one from another.

\subsubsection{Orbital eccentricity effects \\ under the influence of a polar wind}


\begin{figure}[t]
\hspace{-.90cm}
\includegraphics[angle=-90,scale=0.38]{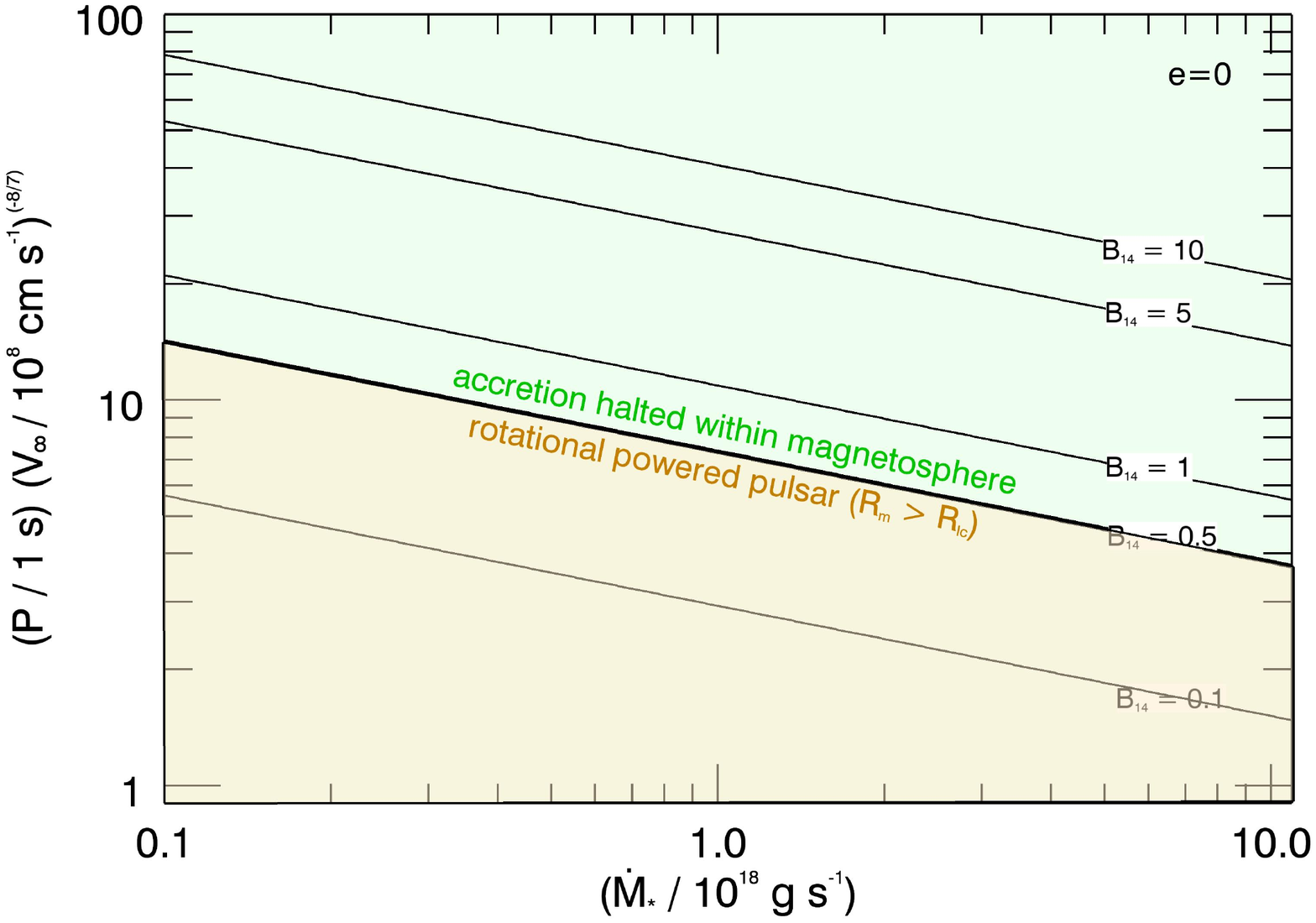}
\vspace{-0.5cm}
\caption{Phase space conditions for accretion regimes around a highly magnetic neutron star under the influence of a polar wind of terminal velocity $V_
\infty$ and mass-loss rate $\dot M_*$. It is assumed that the radius of the neutron star is 10$^6$ cm and its mass is 1.4 M$_\odot$. An example of the phase space separation is marked for the case of
surface magnetic field of 5 $ \times 10^{13}$ G. The eccentricity is assumed equal to zero and
the \lsi\ semi-major axis is adopted in this example.}
\label{11}
\end{figure}

\begin{figure}[t]
\hspace{-1cm}
\includegraphics[angle=-90,scale=0.38]{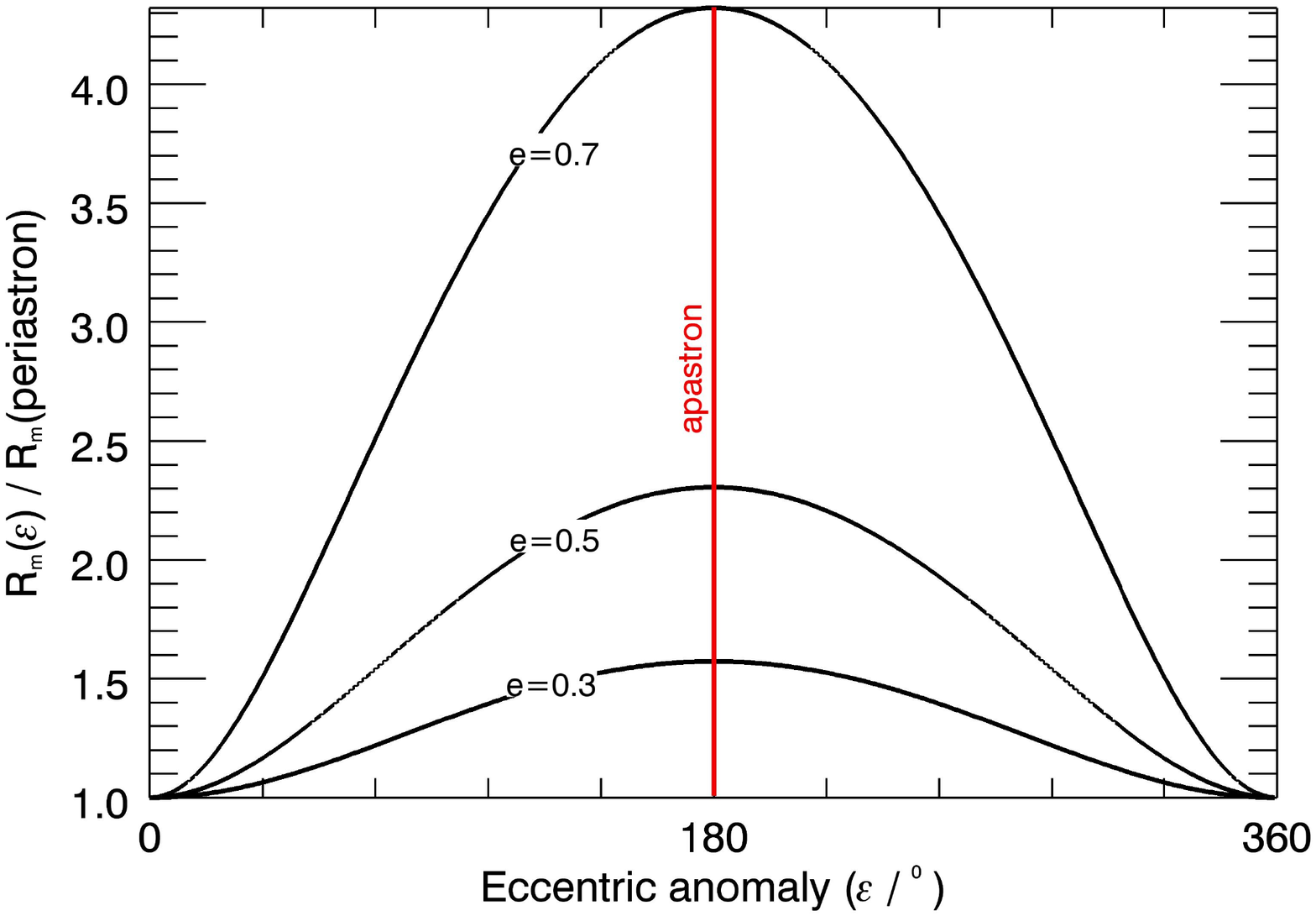}
\vspace{-0.5cm}
\caption{Ratio of the $R_m$-values along an eccentric orbit (under the influence of a polar wind of terminal velocity $V_
\infty$) with respect to that attained at periastron, for different eccentricity, as a function of the eccentric anomaly.}
\label{22}
\end{figure}


Fig. \ref{11} shows 
the condition $R_m = R_{lc}$ as given by Eq. (\ref{Rm=Rlc}) for different values of the neutron star (surface) magnetic field. We adopt one order of magnitude up and one down for the possible 
variation in these parameters out of their fiducial values noted in Eq. (\ref{Rm=Rlc}). 
We assume that the radius of the neutron star is 10$^6$ cm and its mass is 1.4 M$_\odot$.  We adopt a semi-major axis value
of $6 \times 10^{12}$ cm and a radius of the massive star equal to 10 R$_\odot$, consistent with measurements of the 
\lsi\ system (see e.g., Casares et al. 2005, Grundstrom et al. 2007).

An example of the different
physical behaviors of the system is noted in Fig. \ref{11} by the line with $B=5 \times 10^{13}$ G, separating the region of the plane, for that specific value of magnetic field, where the accretion is halted within the magnetosphere (in light green, where $R_m < R_{lc}$),  from that in which the systems acts as a rotational powered system (in light yellow, where $R_m > R_{lc}$). This separation 
is valid at each of the possible (neutron star surface, dipolar)  magnetic field strengths, which form a continuum throughout the plot. 

Fig. \ref{11} and Eq. (\ref{Rm=Rlc})
show, as mentioned in the last Section, that the observed values for periods  and magnetic fields of known magnetars, together with fiducial (and commonly adopted) values
of the stellar (polar) wind velocity and mass-loss rate of \lsi\ would make, assuming a circular orbit, $R_m \sim R_{lc}$. For example, a system hosting a pulsar with
$P\sim 7$ s, where measured periods of magnetars cluster, and the fiducial values for the properties of the wind of the massive star, $V_w=10^8$ cm s$^{-1}$ and $\dot M_*  \sim 10^{18}$ g s$^{-1}$, would be right at the line representing 
$R_m = R_{lc}$ for a surface magnetic field 
$B=5 \times 10^{13}$~G. 

\begin{figure}[t]
\hspace{-.90cm}
\includegraphics[angle=-90,scale=0.38]{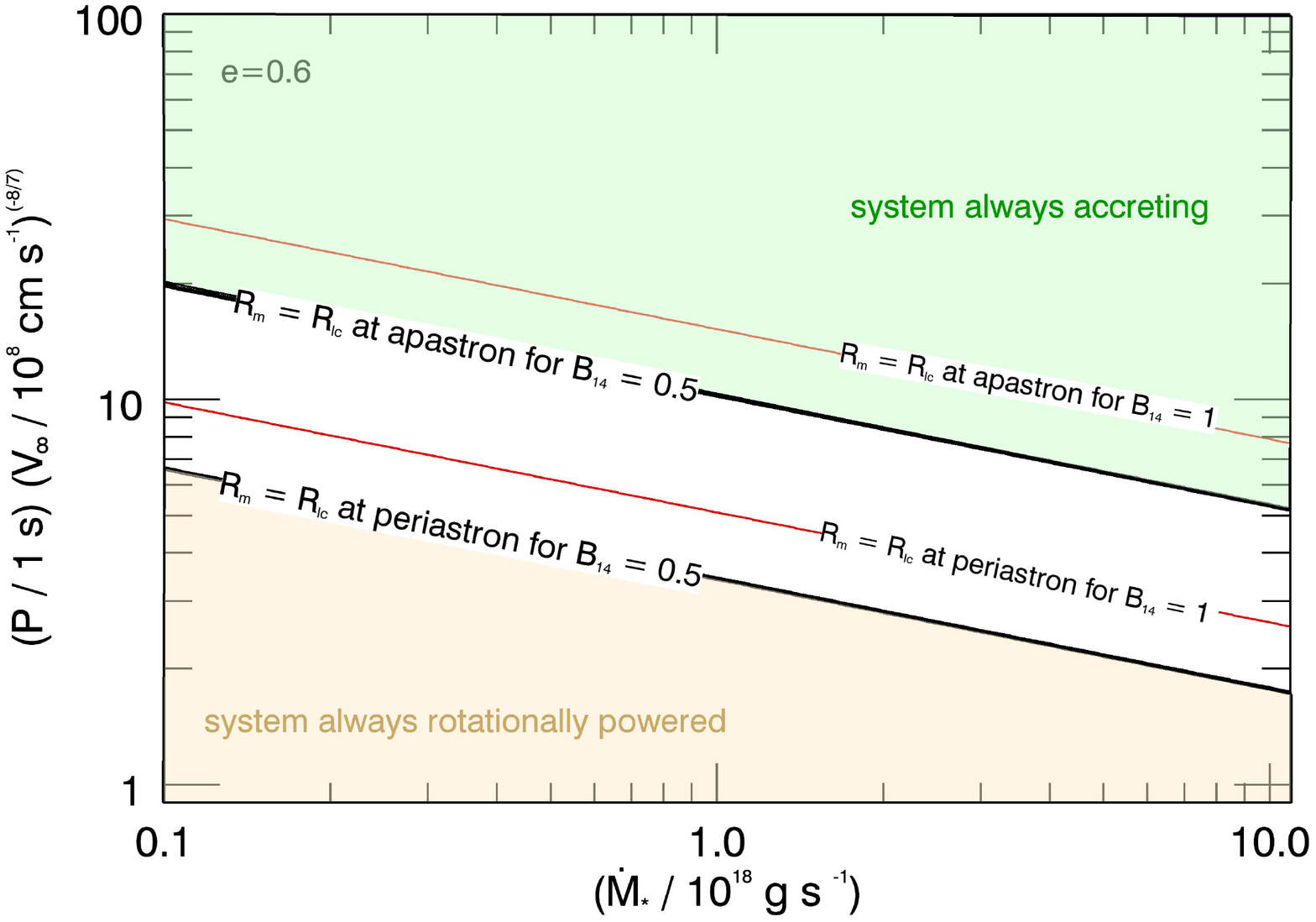}
\vspace{-0.5cm}
\caption{ 
Examples of the region of the phase space 
for a system similar to \lsi\ where the system would change behavior from accreting within the magnetosphere (in periastron) to be a rotationally powered pulsar (in apastron). Each of these phase space regions producing this flip-flopping behavior are located within the two lines
corresponding to equal magnetic field label; for which several pairs are shown. An example with $B=5 \times 10^{13}$~G is explicitly depicted (the central white region of the plot is the
flip-flopping area; the top green part corresponds to always-accreting systems, and the bottom light-yellow one to neutron stars always acting as a rotational power pulsar along the orbit). The flip-flopping region moves up in the plot for higher magnetic fields (an example with $B=10^{14}$~G is shown with red lines).
The neutron star is assumed to be subject to the influence of a polar wind of terminal velocity $V_
\infty$. }
\vspace{0.5cm}
\label{33}
\end{figure}

However, one has to take into account that the orbit of \lsi\ is not circular, and its eccentricity have been quoted in the range of 0.55 to 0.72 (again see, e.g., Casares et al. 2005, Grundstrom et al. 2007, Aragona et al. 2009). 
$R_m$ then becomes a function of the  orbital position, and can be represented, for a given value of eccentricity, as a function of the system's eccentric anomaly.
Fig. \ref{22} shows that the ratio of the $R_m$-values  along an eccentric orbit with that attained by $R_m$ at periastron. For the quoted eccentricity, 
$R_m$ is a factor between 2 and 4 smaller at periastron than what it is at apastron.
Then, if the magnetar-composed system
fulfills the condition $R_m > R_{lc}$ at apastron and the inner magnetosphere thus behaves as that of a rotational powered pulsar, 
it will likely fulfill $R_m < R_{lc}$ at periastron. 
This is graphically shown in Fig. \ref{33}, which shows, for $e=0.6$, the region of the phase space in which a
magnetar hosted in \lsi\ would change behavior. Fiducial values of all parameters, under the assumption of a dominant polar wind, 
would position the system right
in the middle of the flip-flop regime, with the putative magnetar in \lsi\ acting as a rotational powered system in apastron ($R_m$ is larger than the light cylinder in apastron), and accreting within the magnetosphere in periastron ($R_m$ is smaller than the light cylinder in periastron). For usual magnetar parameters, this is the natural solution for the \lsi\ behavior along its orbit.

\subsubsection{Kind of accretion under the influence of a polar wind}

In order to consider further the possible accretion scenario, we will define two additional radii. Let us consider first 
the rotational velocity of the magnetosphere at the position $R_m$, which is given by
\be
V_{rot} = \frac {2 \pi R_{m}} {P},
\label{VVrot}
\ee
and the Keplerian velocity of the infalling matter, which is instead given by
\be
V_{Kep} = \sqrt { \frac{GM_{ns} } {R_m} }. 
\ee
These two velocities allow for the definition of the co-rotation radius, $R_{cor}$,
when $V_{Kep} = V_{rot}$, which can be written as
\ba
R_{cor} &=& \left(  \frac{G M_{ns} P^2}{4 \pi^2} \right)^{1/3} \nonumber \\
&=& 1.67 \times 10^{8} \left( \frac {M_{ns}} {1.4M_\odot} \right)^{1/3}
\left( \frac {P} {1\; {\rm s}} \right)^{2/3} {\rm cm}.
\label{Rcor}
\ea
The co-rotation radius represent the position of the centrifugal barrier for the in-falling material created by the neutron star rotation.  
$R_m > R_{cor}$ when 
\ba
   \left( \frac{P}{ 1\; s} \right) &<& 1456.3 \left ( \frac{ B} {10^{14} \; G } \right)^{6/7}
\left ( \frac{\dot M_* } {10^{18} {\rm \; g \; s^{-1}} } \right)^{-3/7} 
\nonumber \\
&& \times
\left ( \frac{ V_\infty} {10^8 {\rm \; cm \; s^{-1}} } \right)^{12/7} \!\! \left ( \frac{a } {10^{12} \; {\rm cm}} \right)^{6/7} \!\!
 \left( \frac {R_{ns}} {10^6 \; {\rm cm}} \right)^{18/7} \nonumber \\
 && \times   \left( \frac {M_{ns}} {1.4 M_\odot} \right)^{-37/14} (1 -e \cos(\epsilon))^{6/7} 
\nonumber  \\
         & &     \times  \left( 1- 0.69 \left(\frac{ R_*}{ 10 R_\odot } \right)
         \left(\frac{ a } {10^{12} \; {\rm cm} }\right)^{-1} 
         \right.
         \nonumber \\
         && \hspace{2cm} 
         \left.
         \left(1- e \cos(\epsilon) \right)^{-1}  \right)^{12\beta/7} \!\!.
         \label{resp}
\ea
That is, unless values far-off from the fiducial ones are invoked, the inequality in Eq. (\ref{resp}) is always respected. Particularly, when
the period is such that $R_m$ is less, but of the same order than $R_{lc}$, the system is halting accretion at distances from the neutron star which always results in
super-Keplerian velocities.

\begin{figure}[t]
\hspace{-.90cm}
\includegraphics[angle=-90,scale=0.38]{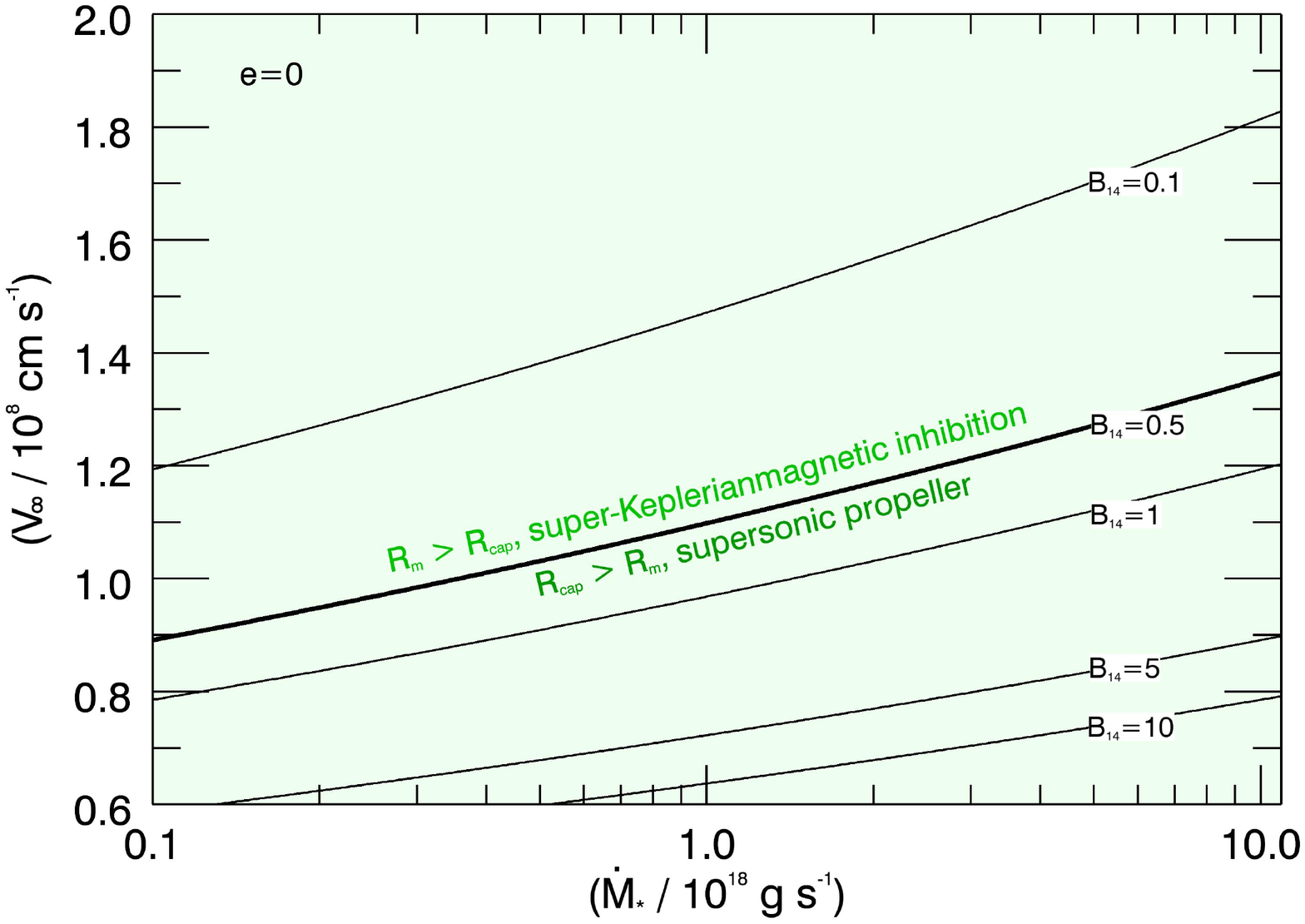}
\vspace{-0.5cm}
\caption{Phase space conditions for accretion regimes around a highly magnetic neutron star. Above each of the curves, corresponding to different values
of the magnetic field (an example is given for $B=5 \times 10^{13}$ G) the system is a super-Keplerian magnetic inhibitor. Below this line,  the system acts as a supersonic propeller. The eccentricity of the orbit is assumed as zero in this plot. The light-green shadow stands for the fact that the system halts matter within the magnetosphere in the whole of the phase space depicted.}
\vspace{0.5cm}
\label{vel}
\end{figure}

\begin{figure}[t]
\hspace{-.90cm}
\includegraphics[angle=-90,scale=0.38]{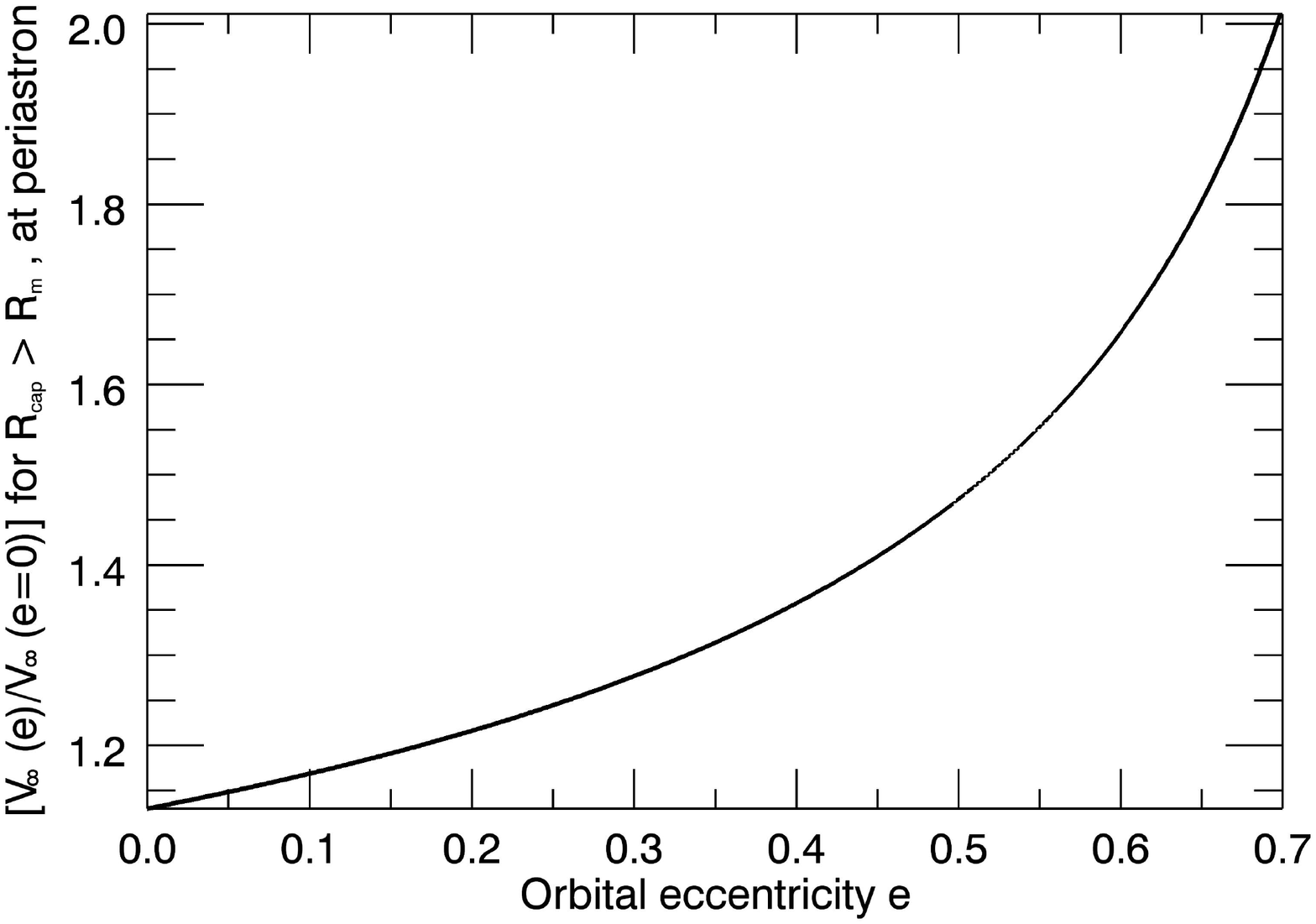}
\vspace{-0.5cm}
\caption{Ratio of the maximal $V_\infty$ value below which the system is a propeller,
for a given eccentricity, with respect to the needed value in an circular orbit. The higher the eccentricity the larger is the value 
of $V_\infty$  that can accommodate a propeller behavior for a highly magnetic neutron star. }
\vspace{0.5cm}
\label{vel2}
\end{figure}

When the accretion is halted in the magnetosphere, i.e., $R_m < R_{lc}$, at a position which is super-Keplerian, 
$R_m > R_{cor}$, the system can be in either magnetic inhibition regime when
$R_{cap} < R_m$, 
or in the stage of supersonic propeller when 
$R_{cap} > R_m$ (see, e.g., Bozzo et al. 2008). 
To decide between these latter possibilities it is useful to consider the relative extent of $R_{cap}$ with respect to $R_{lc}$.
Indeed, when we already know that the matter proceeds within the magnetosphere, i.e., $R_m < R_{lc}$, if $R_{lc} < R_{cap}$,
we also have that $R_m < R_{cap}$ and the system behaves as a supersonic propeller. The inequality
$R_{lc} < R_{cap}$ happens when the period of the neutron star fulfills the following relation,
\ba
\left( \frac{P}{ 1 \; {\rm s} }\right)  &<&
 7.87  \left ( \frac{ V_\infty} {10^8 {\rm \; cm \; s^{-1}} } \right)^{-2}    \left( \frac {M_{ns}} {1.4 M_\odot} \right) \nonumber \\ && \times
 \left( 1- 0.69 \left(\frac{ R_*}{ 10 R_\odot } \right) 
         \left(\frac{ a } {10^{12} \; {\rm cm} }\right)^{-1}  \right.
         \nonumber \\
         && \hspace{2cm} 
         \left.  \left(1- e \cos(\epsilon) \right)^{-1}  \right)^{-2\beta} \!\!.
\ea
In general though, 
$R_{cap} > R_m$ implies, e.g.,  a condition upon the stellar wind velocity as a function of the other parameters in phase space,
\ba
\left( \frac{ V_\infty} {10^8 {\rm \; cm \; s^{-1} }} \right)  <  &&
1.2
\left ( \frac{ B} {10^{14} \; G } \right)^{-2/11}
\left ( \frac{\dot M_* } {10^{18} {\rm \; g \; s^{-1}} } \right)^{1/11}  \nonumber  \\
&& \times 
 \left ( \frac{a } {10^{12} \; {\rm cm}} \right)^{-2/11} 
 \left( \frac {R_{ns}} {10^6 \; {\rm cm}} \right)^{-6/11}   \nonumber  \\
  &&    \times 
         \left( \frac {M_{ns}} {1.4 M_\odot} \right)^{6/11} 
         (1 -e \cos(\epsilon))^{-2/11} 
\nonumber  \\
         & &     \times  \left( 1- 0.69 \left(\frac{ R_*}{ 10 R_\odot } \right)
         \left(\frac{ a } {10^{12} \; {\rm cm} }\right)^{-1}  \right.
         \nonumber \\
         && \hspace{2cm} 
         \left.\left(1- e \cos(\epsilon) \right)^{-1}  \right)^{-\beta} \!\!.
\ea
Fig. \ref{vel} shows the condition upon $V_\infty$ for different values of the phase space parameters, and  for a circular orbit (zero eccentricity). Several examples are marked (for different values of the magnetic field), represented by the solid lines. Each of these lines separate the behavior of the system from being a propeller (below it) to being 
magnetically inhibited (above it). We particularly note this division for $B=5 \times 10^{13}$ G.
Again, the eccentricity value of the orbit is of  importance. Fig. \ref{vel2} shows the ratio of the maximal $V_\infty$ value below which the system is a propeller
for a given eccentricity, with respect to the value it attains in a circular orbit. For instance, for $e=0.6$  the limiting velocity
is enlarged by a factor 1.65, meaning that each of the curves in Fig. \ref{vel} should be displaced in the y-axis by this factor. For the noted example with 
$B=5 \times 10^{13}$ G and the fiducial value of $\dot M_* = 10^{18}$ g s$^{-1}$, all systems with $V_\infty \simeq 2 \times 10^{8}$ cm s$^{-1}$ or less are supersonic propellers.


\subsection{Describing an equatorial disc outflow
and caveats}

We will now consider the situation in which a neutron star feels the influence of an equatorial disc outflow, the matter in which 
is moving with velocity vector $\bar V_{w,eq} $. The uncertainties in describing the equatorial disc have a much
more pronounced impact in the derivations. This section intends to give some of the details about this,  and, if nothing else, make the caveats in the usual assumptions made in doing analytical treatments more explicit. 

\subsubsection{A Bondi-Hoyle approach and its problems}

We start by coming back to Eq.
(\ref{Rcap-def}) for the $R_{cap}$ definition and first 
study which are the components that intervene in this scenario to give rise to the value of $V_{rel}$. Once $V_{rel}$
is determined, most of the (analytical) literature in the topic continue to assume that the accretion rate is given by
\be
\dot M_{acc} = \pi \frac{ (2GM_{ns})^2 }{V_{rel}^3} \rho_{eq} = \pi R_{cap}^2 \, {V_{rel}} \, \rho_{eq},
\ee
where $\rho_{eq}$ is obtained from the continuity equation for matter radially outflowing, with velocity $V_{eq,r}$, a disc of half-opening angle $\Theta$,
\be
\dot M_*^{eq} = 4 \pi r^2 \sin(\Theta) \; V_{eq,r} \, \rho_{eq},
\label{123}
\ee
and where $\dot M_*^{eq} $ is the mass loss rate in this outflow. These equations already imply that Bondi-Hoyle approximation is a valid one for the accretion process onto the star resulting from the equatorial disc outflow, i.e., that the material entering a cylinder of radius equal to the accretion radius will be captured. 
This may not be true by several reasons. First, because even when the Bondi-Hoyle could be assumed, the accretion radius can exceed the vertical size of the disc, and if so, the real accreted mass should be smaller (see, e.g., Zamanov 1995, Zamanov et al. 2001, and the discussion below). 
Also, short-lived Roche-Lobe overflows can occur in close binaries. This approximation also assumes that the Be disc is not affected in any way by the passage of the compact object, which is wrong based on simulations by Okazaki et al. (2002). In general, 
Be disks in binaries are tidally truncated (Okazaki \&
Negueruela 2001) and because of it, it is 
very hard to estimate the accretion rate onto the compact companion without running numerical simulations (see, e.g., 
Okazaki et al. 2002 and 
Romero et al. 2007).

It is nevertheless instructive to see how large an accretion rate one gets by using the Bondi-Hoyle approximation,
and particularly, how much it is sensitive to changes in the assumptions made.
Regarding the other components of $V_{rel}$ apart those contained in 
$\bar V_{w,eq} $,
 from Kepler's law the orbital velocity of the neutron star around the star is given by, 
\be
V_{orb} = \sqrt {    \frac{GM_*^3}{ (M_* + M_{ns})^2} \left( \frac {2}{d} - \frac{1}{a} \right) },
\ee
where $M_*$ is the mass of the stellar companion and the remaining magnitudes have been already defined. 
In terms of the true anomaly $\theta$, defined with the convention of having $\theta = 0$ at periastron, we can write
\be
d = \frac{a (1-e^2)}{1+e \cos(\theta)},
\ee
for the neutron star - Be star separation. 
As we will see, the magnitudes for the forthcoming assumptions for the equatorial wind velocity 
make $V_{orb}$ no longer negligible with respect to the other components of $V_{rel}$.
We have already used that the equatorial disc matter will be assumed to move with a radial, 
$V_{eq, r}$ velocity. In addition it may have also an azimuthal, 
$V_{eq, \phi}$, component --assumed positive for rotation in the direction of the orbital motion--
such that $V_{w,eq}^2 = V_{eq, r}^2 +   V_{eq, \phi}^2 $. Once these are defined, 
the relative velocity is given by $\bar V_{rel} = \bar V_{w,eq} - \bar V_{orb}  $ and its magnitude is given by
\ba
V_{rel} ^2  =& & V_{orb}^2 + V_{eq, \phi}^2 + V_{eq, r}^2 \nonumber \\
 &&
- 2 V_{orb} V_{eq, \phi} \cos (\phi) - 2 V_{orb} V_{eq, r} \sin (\phi),
\label{vrel}
\ea
where $\phi$ is the flight-path angle between the neutron star velocity and the local horizon; measured from the latter to
the neutron star velocity vector. The 
flight-path angle is positive when the neutron star is traveling from periastron to apastron, and negative otherwise.

\begin{figure}
\centerline{\includegraphics[height=7.2cm]{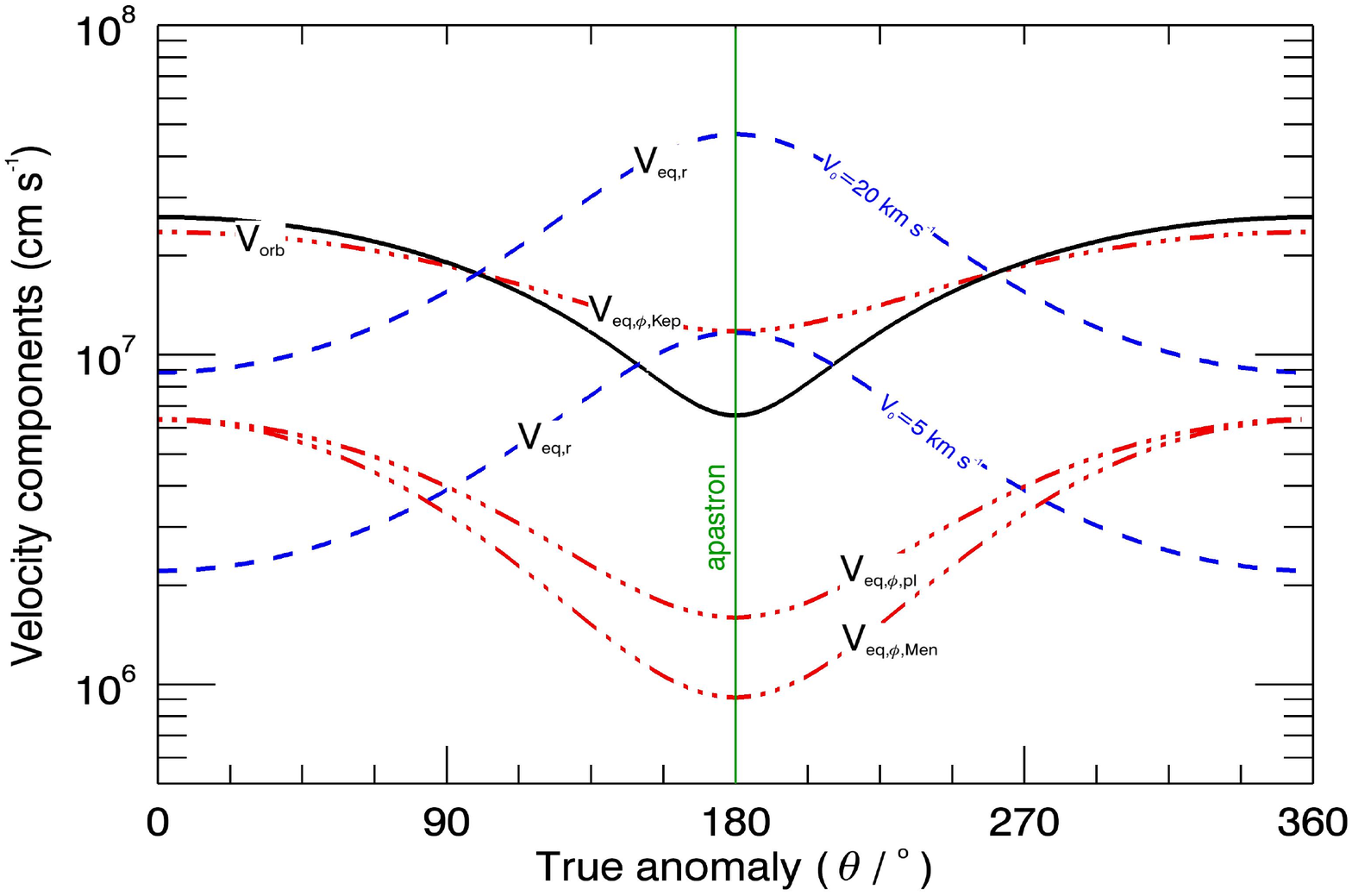}}
\caption{Components of the velocity vectors considered. In red (three-dot dashed) we plot alternatives for the assumption of the
circular velocity of the wind matter. In blue (dashed) we plot a couple of radial velocity components with different pre-factor. The orbital speed of the neutron star is depicted by the solid-black line. }
\label{velcom}
\end{figure}

To proceed further, we need to define the functional form adopted for $V_{eq, r}$ and $V_{eq, \phi}$. 
However, the magnitude of those is far from clear, as is the case, for instance, also for 
the termination distance of the equatorial wind, or its granularity. In this exploratory Section, we can only base our study in different analytical formulations that have been already used in the literature to represent these components. 
The first obvious assumption is to consider that there is no azimuthal movement in the matter of the equatorial disc, $V_{eq, \phi , \, 0}=0$ (e.g. Zamanov et al. 2001, Gregory and Neish 2002).
Gregory and Neish (2002) considered several other parameterizations for $V_{eq, \phi}$, that go from a Keplerian velocity
$V_{eq, \phi , \, {\rm Kep}} = \sqrt{G M_* / r}  $; which would be the natural choice for a viscous decretion disc, 
to different power-laws assumed for the rotation of the envelope
$V_{eq, \phi , \, {\rm Men}} = V_* (r/R_*)^{-1.4}$ (see also, Mennickent et al. 1994) where $V_*$ is the stellar rotation velocity --set to 360 km s$^{-1}$, or $V_{eq, \phi , \, {\rm pl}} = V_{0, \phi}  (R_* / r) / \sin(i)$ with $i$ the inclination and $V_{0, \phi} \sim 1$ km s$^{-1}$ (see also Casares et al. 2005 and Bosch-Ramon et al. 2006). As can be seen in Fig. \ref{velcom} these latter alternatives produce similar (within an order of magnitude) values for the magnitude of the azimuthal velocity component along the orbit, all of them also similar to the orbital speed.

To define the radial velocity component one can also assume, to first approximation, that it is close to zero. Outflow velocities are so low that typically only upper limits (of the order of 1 km s$^{-1}$) are obtained observationally (Okazaki 2010). Marlborough et al. (1997) used for instance $V_0$=0.3 km s$^{-1}$ for their disc parameters.
One could also 
adopt the power-law wind density $\rho_{eq} \propto r^{-n}$, with $n=3.2$, used by Waters (1986) and Mart\'i \& Paredes (1995) in fitting their near-infrared data. For a constant outflow rate, the continuity equation leads to a radial outflow of the form $V_{eq, r} = V_0 (r / R_s)^{n-2}$, where again $r$ is the distance measured from the companion star, and $V_0$ is quoted from a few to a few tens of km s$^{-1}$. $V_0$ was assumed as 5 km s$^{-1}$ in Waters (1986) and in the range 2 -- 20 km s$^{-1}$ in
Mart\'i \& Paredes (1995); what fixes the density of the equatorial wind matter at the star surface on the order of $10^{-11}$ g cm$^{-3}$, with a half-opening angle $\Theta=15^o$. Gregory \& Neish (2002) obtain the radial velocity 
starting from the azimuthal one, forcing the accreted mass to be related to the radio emission along the orbit. The radial velocities obtained in this manner are, for distances smaller than 10 $R_s$, quite similar to the functional form assumed above (see figure 5 in 
Gregory \& Neish's paper). In addition,  we take into account 
that the differences beyond this distance will not play a relevant role if the disc truncates, what is modeled 
with a phenomenological cut to the density of the equatorial wind beyond a distance of $\sim 10 R_*$ (see, e.g., Gregory \& Neish 2002, Bosch-Ramon et al. 2006). By assuming a cut at $\sim 12 R_*$ the second accretion peak seen in some of the curves below is also preserved. 


\begin{figure*}
\hspace{-0.8cm}
\hbox{
\includegraphics[height=7.5cm]{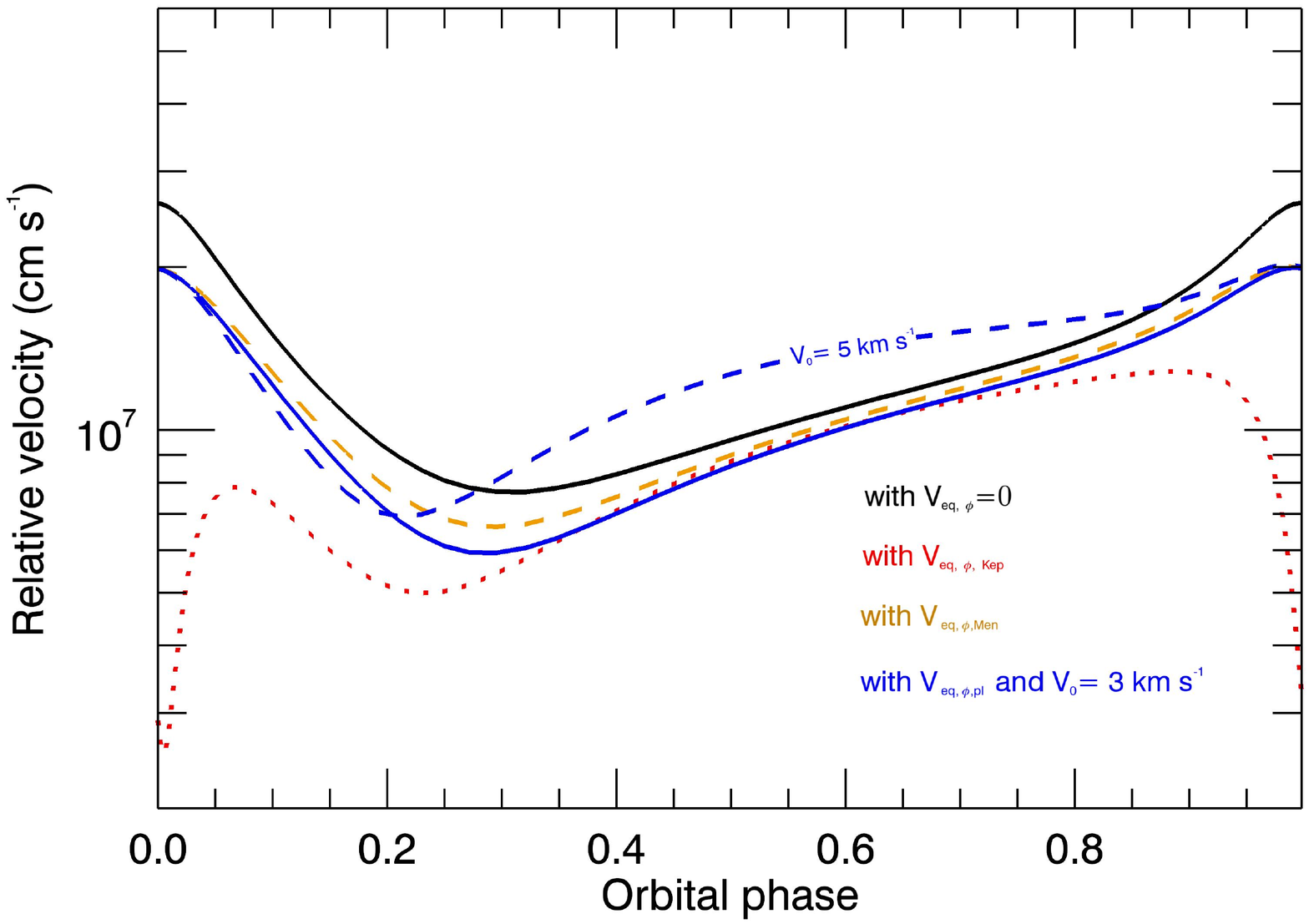}
\hspace{-1.5cm}
\includegraphics[height=7.5cm]{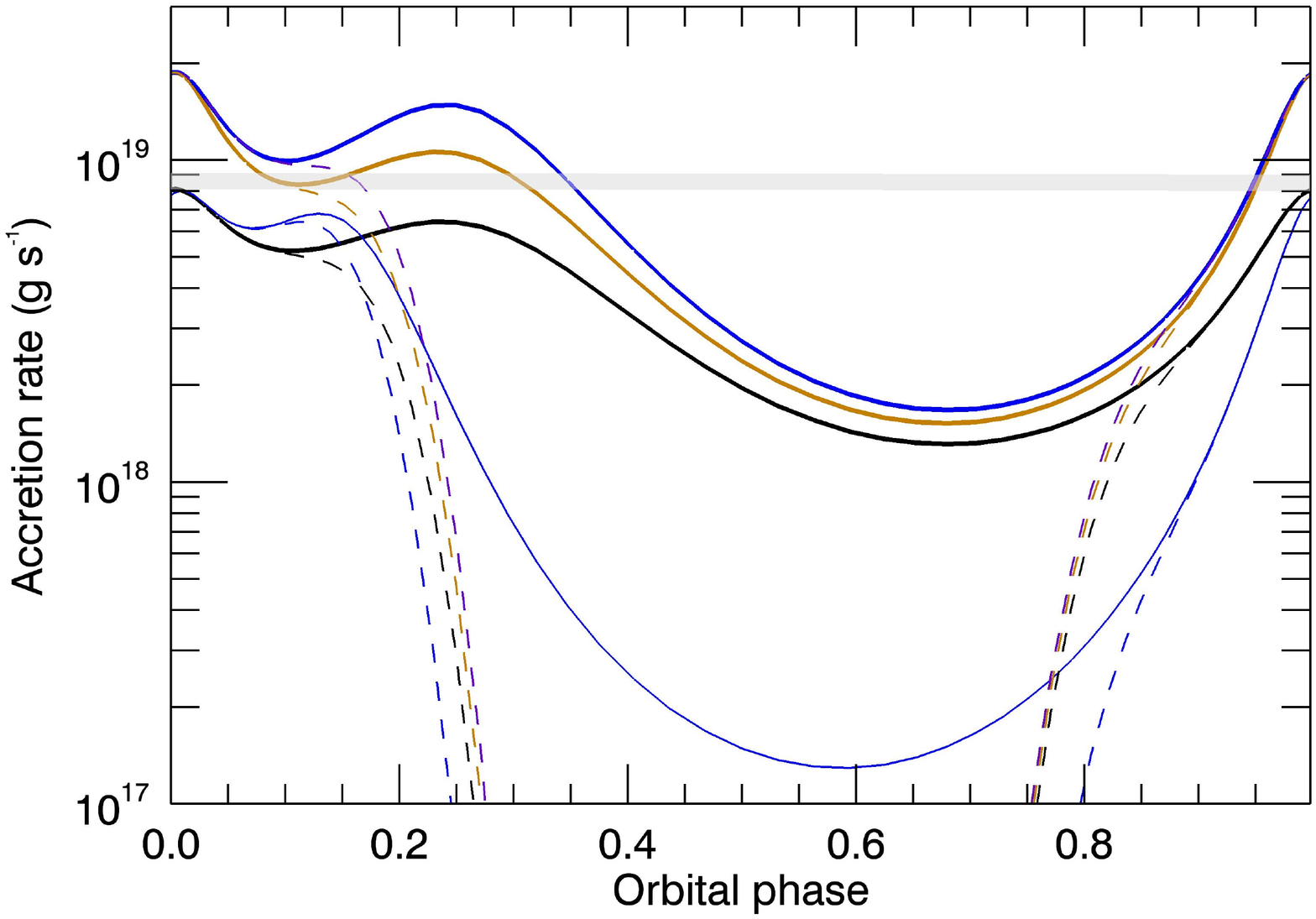}}
\vspace{-0.5cm}
\caption{{\em Left panel}. Relative velocity along the \lsi\ orbit (phase zero is the system's periastron)
for different alternatives for the description of the equatorial wind outflow. No density cut is assumed in this plot. The blue lines
correspond to assume $V_{eq, \phi , \, {\rm pl}} = 1.17 \times 10^{7}  ( (R_* / r) / \sin(i))$ cm s$^{-1}$, for different assumptions of the $V_0$-factor in $V_{eq, r}$. {\em Right panel}. Accretion rate onto the neutron star obtained by means of the Bondi-Hoyle formulation for different relative velocities.
The color coding follows Fig. \ref{vrelplot}, see text for details.}
\label{vrelplot}
\label{mdoteq}
\end{figure*}

In Fig. \ref{vrelplot} we plot $V_{rel}$
as a function of the orbital phase --defined in terms of the eccentric anomaly
as $\phi_p = (\epsilon - e  \sin(\epsilon))/( 2\pi) $, so that the phase is equal to zero at periastron-- for the different alternative descriptions of the outflow previously considered. In particular, we see that large differences appear along the orbital phases of the \lsi\ system for the Keplerian's, Mennickent's, and power-law descriptions of the azimuthal velocity. For the latter case, this plot shows also how large is the influence of a changing initial velocity in $V_{eq, r}$, from 3 to 5 km s$^{-1}$ along the whole orbit. For completeness we also show the case of zero azimuthal velocity (the case of zero radial velocity would produce a similar curve to the Keplerian case, depticted in red in the figure, particularly at low phases).
The accretion rate is very sensitive to small changes in $V_{rel}$. 

Fig. \ref{mdoteq} also shows the accretion rate onto the neutron star, obtained via Eq. (\ref{123}), 
for a few of the cases.
The solid lines stand for the results corresponding to the non-dissolving equatorial disc; whereas the dashed lines correspond to those in which the disc density has been cut off beyond 10 $R_s$. The two blue lines correspond to the result with an initial radial velocity of 3  (top curve) and 5 km s$^{-1}$ (bottom curve), respectively. The horizontal line stands for the stellar mass-loss rate assumed (1.3$\times 10^{-7}$ M$_\odot$ yr$^{-1}$), consistent with the equatorial disc density at the surface being ($\sim 1 \times 10^{-11}$ g cm$^{-3}$). 
The fact that some of the curves are on top of the horizontal line in Fig. \ref{mdoteq} 
must be understood as an unphysical effect produced by the Bondi-Hoyle formula. This situation was also found earlier
(see, e.g., Zamanov 1995, Mart\'i \& Paredes 1995, Bosch-Ramon et al. 2006, Orellana \& Romero 2007, etc.) albeit the plotting of the ratio of the accretion rate with respect to the value it attains at periastron, might perhaps make this fact less clear at first sight. The failure of the Bondi-Hoyle formulae was treated differently by different authors, from ignoring it, given the approximate character of the studies (see above-quoted references), to the addition of one or several ad-hoc cuts (Zamanov et al. 2001), to
the use of radio data to re-scale it. In the last case, 
Gregory \& Neish (2002) assumed that the accretion rate value should be corrected by making 
$
S(r) = K(r) \dot M_{acc} ,
$
where $S(r)$ is the radio flux density (as measured along the orbit at e.g., 8.3 GHz)  and $K(r)$ is the factor of proportionality, depending on the objects separation or orbital phase. 
The former expression can be re-written
as 
$
k(r) \rho(r) = S(r) V_{rel}^3,
$
where $k(r) = K(r) \pi (2GM)^2$. If $V_{rel}$ is computed, $S(r)$ is taken from data, and $\rho(r)$ is assumed 
equal to $\rho_{eq} \propto r^{-n}$ as above,  $k(r)$ can be derived and $K(r)$ immediately follows (Gregory \& Neish 2002, see their figure 6). From this approach all values of the accretion rate are diminished significantly, and correspond to the range from 0.001--0.01 Eddington's; which is equivalent to a reduction of $>$100 times the value of the accretion rate at periastron. 
%
This order of magnitude for the accretion rate would be consistent with the one numerically obtained by Romero et al. (2007). As noted by Orellana et al. (2007, see their figure 1), some of the accretion rate curves of Fig. \ref{mdoteq}
display similar features --e.g., the position of the local maxima in phase--
with the one numerically obtained, but there are approximately three orders of magnitude of difference in 
absolute value.
We note that, consistently, 
Romanova et al. (2003) and Toropina et al. (2006) found 
that the fraction of the Bondi accretion rate which accretes to the surface of the star, obtained from 
MHD simulations
of the accretion to a rotating star in the propeller regime, is much less than Bondi's. We come back to comment on these simulations below.

\subsubsection{Regimes and orbital eccentricity effects \\ under the influence of an equatorial disc outflow}

The analysis in the previous section emphasizes  that the problem of matter exchange in an equatorial wind setting is hard to quantify in an analytical treatment. It can likely be only assessed via numerical simulations, albeit these do not lack their own complexity and caveats (see, e.g., Zdziarski et al. 2010 and Okazaki et al. 2011  for an assessment). Nevertheless, the discussion made in the last section is enough for the point we want to make here: at periastron distances and even considering a significant reduction of the accretion rate shown in Fig. \ref{mdoteq}, if the neutron star pass through the disc
it is likely that accretion from the equatorial outflow  dominates over that from the polar wind, implying $\dot M_{acc, eq} > \dot M_{acc, polar}$. For comparison, we recall that the polar outflow accretion, computed by use of Eqs. (\ref{BHL}) and (\ref{Rcappolar}), lead to values that are maximal at periastron, where it attains $\sim 2 \times 10^{14}$ g s$^{-1}$ (this value is  two orders of magnitude less than the expected equatorial-outflow accreted-matter resulting from numerical simulations),  to quickly fall to $5 \times
10^{12}$ g s$^{-1}$.
In this situation, it is even easier for a magnetar to allow matter entering into its magnetosphere, given that $R_m$ is smaller under the equatorial wind influence. Indeed,
\ba
R_m &=& 8.2 \times 10^{9} \left ( \frac{ B} {10^{14} \; G } \right)^{4/7}\left ( \frac{\dot M_{acc,eq} } {10^{16} {\rm \; g \; s^{-1}} } 
\right)^{-2/7} 
\nonumber  \\ &&
\times
  \left( \frac {M_{ns}} {1.4 M_\odot} \right)^{-1/7} 
\left( \frac {R_{ns}} {10^6 \; {\rm cm}} \right)^{12/7} {\rm cm},
\label{rmeq}
\ea
and in order for matter in this outflow to enter within the magnetosphere, 
\ba
\left( \frac{P}{1 \,s}\right) &>  & 1.7 \left ( \frac{ B} {10^{14} \; G } \right)^{4/7}\left ( \frac{\dot M_{acc,eq} } {10^{16} {\rm \; g \; s^{-1}} } 
\right)^{-2/7} 
\nonumber  \\ &&
\times
  \left( \frac {M_{ns}} {1.4 M_\odot} \right)^{-1/7} 
\left( \frac {R_{ns}} {10^6 \; {\rm cm}} \right)^{12/7} .
\label{Peq}
\ea
The larger the accretion rate at periastron, the easier it is for pulsars with 
lower periods to have matter within the magnetosphere, which is reflected, for fiducial values, 
comparing Eqs. (\ref{Peq}) and (\ref{Rm=Rlc}). 

Aditionally, to ask for the inequality $R_m > R_{cor}$ to be fulfilled implies, using Eqs. (\ref{Rcor}) and (\ref{rmeq}),
a condition over the period
\ba
\left( \frac{P}{1 \,s}\right) &<  & 344 \left ( \frac{ B} {10^{14} \; G } \right)^{6/7}\left ( \frac{\dot M_{acc,eq} } {10^{16} {\rm \; g \; s^{-1}} } 
\right)^{-3/7} 
\nonumber  \\ &&
\times
  \left( \frac {M_{ns}} {1.4 M_\odot} \right)^{-5/7} 
\left( \frac {R_{ns}} {10^6 \; {\rm cm}} \right)^{18/7} ,
\label{Peqrcor}
\ea
which is easy to satisfy for all measured values of $P$ of  magnetars. 
Finally, using the definition of $R_{cap}$, the condition $R_{cap} > R_m$
implies
\ba
\left ( \frac{ B} {10^{14} \; G } \right) &<& 14.1 \left ( \frac{\dot M_{acc,eq} } {10^{16} {\rm \; g \; s^{-1}} } 
\right)^{1/2}  \left ( \frac{ V_{rel} } {10^8 {\rm cm \, s^{-1} }} \right)^{-7/2}
\nonumber  \\ &&
\times
  \left( \frac {M_{ns}} {1.4 M_\odot} \right)^{2} 
\left( \frac {R_{ns}} {10^6 \; {\rm cm}} \right)^{-3} .
\label{Beq}
\ea
We lack accurate knowledge of $V_{rel}$ to assess the former inequality, other than observing that 
the fiducial value chosen is --to be conservative-- 
one order of magnitude larger than the one shown in Fig. \ref{vrelplot}. A smaller value for $V_{rel}$ (smaller than the one chosen as fiducial) would make the inequality even easier to fulfill. 
Then, this condition is not restrictive for magnetars, implying that, under the influence of an equatorial
outflow with an accretion rate of the order of $10^{15  - 16}$ g s$^{-1}$, and with the caveat of not having a 
precise description of short-lived effects such as a Roche-Lobe overflow or a formation of a transient accretion disc, 
the system would also act as propeller in its periastron.

\section{Energetics}
\label{ENER}

\subsection{Maximal energy of electrons at periastron shocks}

To analyze the possible maximal electron acceleration in the propeller we consider, as, e.g., Bednarek (2009),
the  rate of energy increase and of energy losses in a shock formed at the magnetic radius position; where turbulent
motion of a strongly magnetized medium is considered prone to particle acceleration. The acceleration gain can be parameterized as 
\be
\left( \frac{dE}{dt } \right)_{acc} = \zeta c E / R_L = \zeta c e B
\ee
where $\zeta$ is an efficiency of acceleration (fiducial value to be taken as 10\%), $R_L$ is the Larmor's radius, 
and $e$ is the electron charge.   $B$ is the magnetic field, which value will be obtained from Eq. (\ref{BB}).
$\zeta$ is a free parameter here and the fiducial value for it has been chosen conservatively;
generally $\zeta \ll 1$, and is close to 1 only in extreme accelerators (e.g., Aharonian et al 2002, 
Khangulyan et al. 2007, 2008, considered values of $\zeta \sim  10^{-2} - 10^{-4}$).

Given that electrons of the highest energies loose energy mostly via  the synchrotron process, 
 \be
\left( \frac{dE}{dt } \right)_{loss} = \frac 43 c \sigma_T \rho_B \gamma^2.
\ee 
Here, $\sigma_T$ is the Thompson cross section, $\gamma$ is the electron Lorentz factor, and $\rho_B=B^2/8\pi$ is the energy density of the magnetic field. The maximum energy of the electrons are determined by the balance of the two former equations, 
which, by means of Eq. (\ref{rmpolar}), results in
\ba
\gamma_{max}^{(polar)}&=&1.1 \times 10^7 \left( \frac{\zeta}{0.1} \right)^{1/2} 
\left ( \frac{ B} {10^{14} \; G } \right)^{5/14}
\nonumber  \\
&&
 \times
 \left ( \frac{\dot M_* } {10^{18} {\rm \; g \; s^{-1}} } \right)^{-3/7} 
 \left( \frac{ V_\infty} {10^8 {\rm \; cm \; s^{-1}} } \right)^{12/7}
\nonumber  \\
         & &
          \times
          \left ( \frac{a } {10^{12} \; {\rm cm}} \right)^{6/7}   \left( \frac {R_{ns}} {10^6 \; {\rm cm}} \right)^{15/14}\nonumber  \\
         & &
             \times     
         \left( \frac {M_{ns}} {1.4 M_\odot} \right)^{-15/14} \! (f(e))^{3/2},
         \label{gm}
\ea
where
\ba
f(e)&&=(1 -e \cos(\epsilon))^{4/7}  \nonumber  \\
         & &     \times  \left( 1- 0.69 \left(\frac{ R_*}{ 10 R_\odot } \right)
         \left(\frac{ a } {10^{12} \; {\rm cm} }\right)^{-1} \right.\nonumber  \\
         && \left.
\hspace{2cm} \left(1- e \cos(\epsilon) \right)^{-1}  \right)^{8\beta/7}.
\ea
In periastron, and with $a=6 \times 10^{12}$ cm, for fiducial values,
the maximal electron energy prefactor (in Eq. (\ref{gm}))
reaches $6.8 \times 10^{12}$ eV. If we instead use the equatorial result derived in Eq. (\ref{rmeq}), 
the maximal energy would read 
\ba
\gamma_{max}^{(eq)}&=&
2.7 \times 10^6 
\left( \frac{\zeta}{0.1} \right)^{1/2} 
\left ( \frac{ B} {10^{14} \; G } \right)^{5/14} 
\nonumber  \\
         & &
    \times     
    \left ( \frac{\dot M_{acc,eq} } {10^{16} {\rm \; g \; s^{-1}} } \right)^{-3/7} 
 \left( \frac {R_{ns}} {10^6 \; {\rm cm}} \right)^{15/14}
 \nonumber  \\
         & &
    \times     \left( \frac {M_{ns}} {1.4 M_\odot} \right)^{-3/14} .
         \label{gm2}
\ea
The maximal energy in this case is then of the order of $1.3 \times 10^{12}$ eV, for fiducial values.
Note that the higher the accretion rate (or the lower the magnetic field), the lower the maximal energy. 
Of course, the smaller the value of the efficiency, $\zeta$, which could also change along the orbit,
the smaller the maximal energy.  To be explicit, 
for an assumed value of $\eta = 0.01$ (e.g., Khangulyan et al. 2008), the maximal energy is reduced to 411 GeV. The turbulent region of the propeller shock would not make electrons to achieve TeV energies.



\section{Discussion}
\label{discussion}

In the previous Sections we have discussed the consequences of having a magnetar in 
the \lsi\ system (or at least a high $B$, long $P$ pulsar able to burst in the way described). 
 This possibility was based:
\begin{itemize}
\item on the observational evidence for one very short flare observed by {\it Swift}, for which the only remaining counterpart candidate X-ray source within an improved 1$\sigma$ uncertainty in position is  \lsi, 
\item on the complete similarity of the previous event with all properties known 
for other magnetar flares,
\item and on the fact that the longer flares observed by other satellites (and undoubtfully coming from \lsi,  having timescales of about 1 h) are uncorrelated with harder X-ray emission, so discarding that the shorter flare is a result of a strong spectral evolution of a typical non-magnetar burst (proven by an uncorrelated {\it Swift}--BAT or {\it RXTE}-HEXTE detection in  any of the longer flares).
\end{itemize}
If we are to accept that a magnetar may be part of the \lsi\ system (which would then be the first such in being discovered) 
we concluded from the analysis in the former Section that it would likely be subject to a flip-flop behavior. It would shift from behaving as a neutron star being rotationally powered near apastron
to being a propeller near periastron, along each of the system's orbit. 

Based on the preliminary analysis of the {\it Swift}--BAT burst
as reported in Astronomer Telegrams (De Pasquale et al. 2008; Barthelmy et al. 2008; Dubus \& Giebles 2008), earlier considerations of \lsi\ being formed by a magnetar have been made (Bednarek 2009, Dubus 2010). 
In them, the system has not been proposed to flip-flop on any behavior, and the influence of the orbit was not considered. Dubus (2010) proposed that the system would always behave as a normal pulsar, mimicking in all respects the properties needed by an inter-wind shock to be sustained (e.g., as in Dubus 2006).  He also proposed that the short burst suggested a magnetic field strength of 10$^{15}$ G, which we do not find justified. 
Instead, Bednarek (2009) proposed that the \lsi\ system is always a propeller, even in apastron. 
We find above that for typical magnetar parameters, neither of these possibilities seems likely to be the case.

The idea of a flip-flop behavior for a neutron star in an eccentric orbit was earlier considered (see, e.g., Illarionov \& Sunyaev 1975,
Gnusareva \& Lipunov 1985, Lipunov 1987, Lipunov et al. 1994, Campana et al. 1995). And actually, a flip-flop  behavior was even proposed for \lsi\ itself (Zamanov 1995, Zamanov et al. 2001) although the idea was quickly abandoned in the literature perhaps because for ms-pulsars, the system would indeed always be rotationally powered, as shown above. In this work, we have thus rekindled some of these ideas but given them an extra edge: prompted by the observational analysis, the \lsi\ pulsar has magnetar parameters and thus the flip-flop behavior is rather an expected outcome of the orbital evolution. 
But may the flip-flop behavior be in qualitative agreement with the remaining multi-wavelength phenomenology of the system?          

\subsection{A flip-flopping magnetar model for \lsi\  in the context of multi-wavelength
observations}

\subsection{TeV range}

The TeV emission of the \lsi\ system has been discovered by {\it MAGIC} (Albert et al. 2006). A claim of periodic recurrence of the TeV emission was made thereafter, with the system showing regular outbursts at TeV energies in a broad-range of phases around 0.65 (radio phases, as in Gregory 2002, where periastron is at phases 0.23--0.3, see, e.g.,  Casares et al. 2005 or Grundstrom et al. 2007),  with  no significant signal elsewhere (Albert et al. 2009). The {\it VERITAS} array did independent observations and soon confirmed the {\it MAGIC} detection (Acciari et al 2008, 2009). 
In addition, the {\it MAGIC} collaboration claimed a correlated X-ray / TeV emission at the position of the outburst (Anderhub et al. 2009) which would emphasize the likely common origin of the radiation in both bands. However, this was based on 60\% coverage of just one orbit, and there was no clue on how general this result really was. 

Neutron star-based models presented to analyze the TeV emission from \lsi, would produce a recursive orbital behavior, unless 
of course significant parameters (e.g., those defining the level of absorption) 
change in an orbit-to-orbit basis for yet unknown reasons 
(see, e.g., Dubus 2006b, 2010; Sierpowska-Bartosik \& Torres 2007, 2008, 2009; Bednarek 2009). The strong dependence on the system's geometry and the orbital position shown by the absorption and emission process (pair production and anisotropic inverse Compton scattering) would produce an orbit-to-orbit recurrent TeV burst. However, the peak of the TeV emission found by early measurements of {\it MAGIC} and {\it VERITAS}, at radio-phase $0.6 - 0.7$, is not strictly near the superior conjunction of the orbit, where the conditions would be more favorable to gamma-ray production.
Additional considerations where made to correct for this displacement in these models. Dubus et al. (2010)
consider the influence of a relatively mild Doppler-boosted emission; assuming that there is an outflow velocity of 0.15--0.33 $c$, consistent with the expected flow speed at the pulsar wind termination shock. Sierpowska-Bartosik \& Torres (2009) assumed that the lightcurve is generated by a variable power in the injection of relativistic particles.

Nevertheless, new observations by {\it VERITAS} (Acciari et al. 2010) discard the regularity of the TeV burst and the X-ray / TeV correlation.
The observing seasons from 2008 to 2010 covered a total of 8 orbits out of 27 available. 
In none of them the system was detected where it was earlier expected, near apastron. Actually, in 7 out of 8 orbits the system was not detected in any phase; which can not be accommodated, as discussed by Acciari et al. (2010), arguing for shorter observation times. The upper limit imposed were 
less than 5\% of the Crab Nebula flux in the same energy range, in contrast to previous observations by both {\it MAGIC} and {\it VERITAS} which detected the source during these phases at 16\% of the Crab Nebula flux. This implies that the system is in a low TeV state where the flux is at least a factor of 3 lower than the one detected in the 2006 observing season.  
Additionally, {\it VERITAS} reported  the system was significantly detected just once,  during observations taken in late 2010, at a phase much closer to periastron passage. 

In the scenario of this paper, the usual appearance of the TeV emission near apastron would be explained under the common lore: the inter-wind shock produced by the interaction of a rotationally powered pulsar and the stellar wind of the star accelerates particles, which in turn emit gamma-rays. The common disappearance of the TeV emission in periastron, instead, is accommodated by the fact that on the one hand, the energy cutoff for particle acceleration in the shock formed in the propeller regime is sub-TeV (and will be less for larger accretion rate, smaller magnetic field, and the expectedly smaller acceleration efficiency out of the fiducial values assumed); 
and, on the other, the cross section for pair production for any TeV particle is maximal. 
Typically, then, the system would appear and disappear where it has been observed along the 2006 to 2009 seasons. 

Contrary to other models, the flip-flopping \lsi\ system could also provide a natural interpretation for the flux reduction or 
the total disappearance of the TeV radiation. If the mass-loss rate (and thus the accretion rate onto the neutron star) increases, the system more quickly abandons its rotationally powered behavior, and if it increases enough, it would not be in such a behavior at any portion of the orbit. The inter-wind shock would not form and abundant TeV particles would not be produced.  We note --see Fig. \ref{33}-- that even a small factor increase in the mass-loss rate at apastron 
can make the system to have a disrupted magnetosphere there as well. (If the mass-loss rate of the equatorial outflow increase, its properties may change significantly, e.g., the truncation radius could be enlarged and even dominate at apastron, also implying that significantly less TeV emission would be produced there.) 

The orbit-to-orbit TeV changes can be understood in equal terms, by directly linking it to the variations of the mass-loss rate in a highly variable behavior of Be stars. This has the obvious caveat of making a particular orbit's TeV lightcurve impossible to foresee, albeit links observations able to constrain the mass-loss rate with those at the highest energies, and allows for a prediction of anti-correlation to be made: the higher the mass-loss rate the lower the TeV emission. Allowing oneself to speculate, detailed observations of this anti-correlation, if confirmed, can impose constraints over the magnetar period.

\subsection{The TeV emission and its apparent correlation with the radio-obtained super-orbital phase }

We now compare the years-long evolution of the system as tracked in radio and TeV energies and look at it from the perspective of the magnetar-composed \lsi\ model.

\begin{figure}[t]
\hspace{-.90cm}
\includegraphics[angle=-90,scale=0.38]{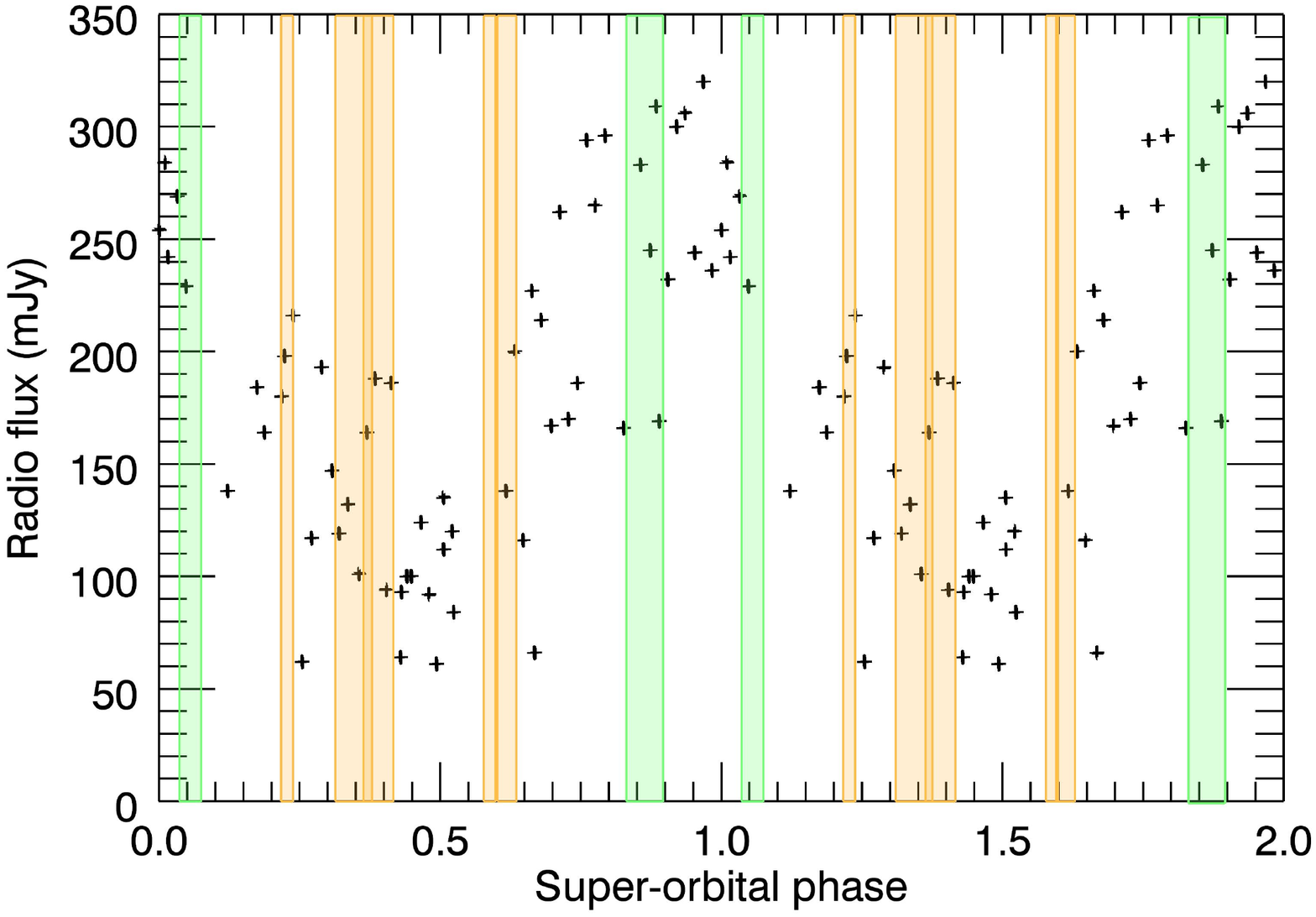}
\vspace{-0.5cm}
\caption{Radio data from \lsi\ compiled by Gregory (1999), most of it at a frequency of 8.3 GHz, and folded with the super-orbital
period of 1667 days. The colored boxes represent {\it MAGIC} and {\it VERITAS} observation times that covered the broadly-defined 
apastron region of the \lsi\ orbit. Light-yellow boxes stand for periods in which the \lsi\ system was detected whereas the light-green boxes stand for those in which it was not, implying at least a factor of 3 reduction in flux. Each of the boxes represent several orbits of \lsi. Two super-orbital cycles are shown for clarity.}
\vspace{0.5cm}
\label{radio-tev}
\end{figure}

For the radio data we use the measurements compiled by Gregory (1999), most of it at a frequency of 8.3 GHz, using the Green Bank Interferometer. The analysis of the radio emission led to the discovery of a super-orbital
period at 1667 days (Gregory 2002). Fig. \ref{radio-tev} shows these data folded with the super-orbital period. Super-imposed, we also show all the super-orbital phase-ranges where there were TeV observations (either by {\it MAGIC} or by {\it VERITAS}) which covered the orbital phases where the \lsi\ system was initially detected near apastron, i.e., at a broad radio phase around $\sim 0.65$. Then, all colored boxes in Fig. \ref{radio-tev} 
represent the TeV observations 
that covered the broadly-defined apastron region of the \lsi\ orbit. 
Typically, several orbits are included in each of the colored boxes.
In order to plot these TeV observations in Fig. \ref{radio-tev}, the MJD of each observation time that covered the apastron region was obtained from all TeV reports on \lsi\ (Albert et al. 2006, 2009; Anderhub et al. 2009; Acciari et al. 2008, 2010) and converted into super-orbital phase using the period of 1667 days. We note that, although sparse, the TeV coverage of the source goes from 2006 to 2010.

From Fig. \ref{radio-tev},
we see that the system has appeared as  a significant TeV-source near apastron only (light yellow boxes)
at super-orbital phases near the minimum of the radio flux. Instead, at the maximum of the radio flux, i.e., at super-orbital phases near $\sim 0$, observations confirmed only  TeV  upper limits (green colored boxes).
The latter situation corresponds to the recent disappearance of the source as reported by {\it VERITAS},  signifying an important flux reduction.
Needless to say, the radio data and the TeV measurements are not contemporaneous, and we will especulate that the average properties of the super-orbital period producing the radio long-term modulation have been maintained since 2002 till now to extract conclusions. This may not necessarily be true (see, e.g., Trushkin \& Nizhelskij 2010), although the drift in the maxima of the radio peak seems very small (of the order of 3 days) to affect the reasoning. 

In itself, it is possible that Fig. \ref{radio-tev} represents a clue to the nature of the source.
If we follow the interpretation by e.g., Gregory et al. (1989),
Zamanov et al (1999), and Gregory (2002) and accept that the super-orbital radio flux modulation is the result of a cyclical variation in the mass-loss of the Be star (the synchrotron radio power emitted is directly proportional to the relativistic
particle density which likely scales with the mass-loss rate),
the disappearance of flux-reduction of the TeV radiation at the peak of the radio emission can be accommodated as follows.
In the magnetar model for  \lsi, the increase in mass-loss rate leads to an increase in the accretion rate onto the compact object, which pressure makes the magnetic radius to become similar and eventually less than the light cylinder, disrupting the magnetosphere even at apastron. If so, the system stops behaving as a normal pulsar, stops driving a wind, and stops forming an inter-wind shock. Instead, \lsi\ acts as a propeller. As a result, it is not expected to produce a significant number of multi-TeV particles, and thus the TeV radiation is suppressed. As in periastron the system was already behaving as a propeller, there is no significant TeV in any part of the orbit.

We note that the results of Fig. \ref{radio-tev} are model-independent, since they are based only on observational results (when there was a detection near apastron, when there was none) and the times in which these observations were taken. They are however difficult to accommodate in other models of the source (which in addition are not able to explain the short burst).
On the one hand, in a microquasar scenario, the increase of the mass-loss rate would enhance --under the disc-jet coupling assumption (e.g., Falcke \& Biermann 1995)-- the power in the jet. Thus, in a microquasar model, when the mass-loss and accretion rate increase there should be more TeV radiation, not less.
On the other hand,
estimations of the variations in the mass loss-rate from the Be star are given as the ratio between maximal and minimal values obtained either from radio emission (a factor of 4 was determined by Gregory \& Neish 2002) or from H$\alpha$ measurements,
which span from a factor of 5.6 (Gregory et al. 1989) to 1.5 (Zamanov et al. 1999); in any case, a factor of a few. In a normal pulsar wind -- stellar wind scenario, when the pulsar is not a magnetar, an increase in the mass loss rate by a factor of a few does not change anything in the behavior of the source (what can be verified using for instance Eq. (\ref{Rm=Rlc})): the system would always have a non-disrupted magnetosphere. In the inter-wind shock scenario with a normal pulsar being part of the system, there is no apparent reason for the disappearance or the significant reduction  of the TeV emission, nor for the apparent correlation of this reduction with the maximum of the mass-loss rate when it increases by a factor of a few.
Instead, in the case of a magnetar in \lsi, fiducial values of the involved magnitudes 
put the neutron star in a position of the phase-space in which small changes in the mass-loss rate can produce regime changes, accommodating Fig. \ref{radio-tev}.

\subsection{GeV range}

GeV emission from \lsi\ has been detected using the Large Area Telescope onboard {\it Fermi} (Abdo et al. 2009), at the  orbital
periodicity. With the data of the first 8 months of the mission, only 
a single spectrum averaged 
along the orbit could be obtained. A power-law with an exponential cutoff at a few GeV was found as the best fit. 
The folded {\it Fermi} lightcurve peaked at periastron passage.
The anti-correlation of the phase of the maximum between TeV and GeV can be understood as a result of 
inverse Compton scattering and pair absorption.
%
With 2.5-years of {\it Fermi}-LAT data more details have been obtained (see, e.g., Torres et al. 2010, Hadasch et al. 2010 for preliminary reports).
The spectral characterization of \lsi\ allows now for a fit of a  
power-law with an exponential cutoff spectrum along each analyzed portion of the systemÕs orbit (which is typically cut in halves). The differences between these spectra is however not large, of the order of 20\% in flux and within errors in the determination of the cutoffs. The LAT now detects emission up to tens of  GeV, where prior datasets led to upper limits only.  Due to the contemporaneous measurements by {\it VERITAS}, commented above, we know now that there is no  evidence that the process responsible for the detected {\it Fermi}-LAT emission continues beyond the GeV cutoff. 
One of the most interesting results of this further {\it Fermi}-LAT campaign is the detection of an overall flux increase around March 2009, of the order of 30\%. This flux change was accompanied by a significant decrease in the GeV lightcurve modulation, which since then is much flatten. This flux change is then approximately coincident with the period of TeV flux reduction.

Within the magnetar-composed flip-flopping model of \lsi\  the usual behavior of GeV emission, being  modulated along the orbit and anti-correlated with the TeV emission, finds its place as commented before. The increase in GeV luminosity like the one found after March 2009 can be understood due to the increase of the accretion rate, which would in turn increase the energy reservoir of which a fraction ends up in the GeV domain. If this is the case, the magnetar in \lsi\ would be behaving as a propeller along more portions of the orbit, and eventually along all of it. One would qualitatively expect that the
GeV emission would not be significantly modulated anymore, since it is the same process generating the radiation all along the orbit (no more inter-wind shock), the GeV modulated fraction would diminish and flatter,  with the TeV emission being simultaneously reduced.

\subsection{X-ray range}

Zhang et al. (2010), Torres et al. (2010) and Li et al. (2011) present the status of our knowledge of the X-ray behavior of the source, via long-term, years-long monitoring using the {\it INTEGRAL} and {\it RXTE} satellites. The latter is the richest dataset, as has been discussed and enlarged via the addition of several more orbits
in Section \ref{OBS} above. Flares in the ks-timescale have been detected since long by different experiments (see references in the quoted papers) and they can be accommodated as local processes, for instance, through clumpiness in the stellar wind. Their interpretation would likely not be affected within the magnetar-composed \lsi\ model.
An outcome of the \rxte\ campaign is the discovery of variability in the orbital profiles of the X-ray emission. Lightcurve  profile variability is seen from orbit-to-orbit all the way up to multi-year timescales, with 
the phase of the profile maximum also changing. Such a high degree of variability can certainly be more easily accommodated with a magnetar hosted by  \lsi\ than in other models. In the former,  we have different processes at different places contributing to the X-ray emission, already in an orbit-to-orbit basis, the predominance of which and the phases at which they become dominant can be altered  by both, random (in orbital timescales) and cyclical (in super-orbital timescales) variations of the accretion rate. 
The possible appearance of the 
super-orbital variability in X-rays remains a matter of study for when additional data accumulates.


\subsection{Radio range}

The existence of periodic ($\sim 26.5$ days) non-thermal radio outbursts have allowed to determine a strong 
modulation in the amplitude occurring on a timescale of $\sim 4$ years, both of which have been commented above (see, e g., Gregory 2002, and references therein). Cyclic variability in the Be star envelope have been proposed as a possible origin of the long term modulation. This has been emphasized by an apparent correlation of the super-orbitally varying radio flux with the H$\alpha$ line parameters; the latter also present super-orbital modulation with the same period, albeit it is shifted in phased about 400 days (e.g., Zamanov \& Mart\'i 2000). 
In terms of the flip-flopping magnetar model for \lsi, the outburst onset could represent the time of the transition of the neutron star from propeller to a normal pulsar behavior (see, e.g,  Zamanov 1995). The discussion in  Zamanov \& Mart\'i (2000) put the general features of this model in context with radio observations (note that the super-orbital period has been corrected from 1584 to 1667 days since that paper was published, and the analysis should be changed accordingly).

Very Long Baseline Array (VLBA) imaging obtained by Dhawan et al. (2006) over a full orbit of \lsi\ has shown the radio emission to come from angular scales smaller than about 7 micro-arcsec (which is a projected size of 14 AU at an assumed distance of 2 kpc). This radio emission appeared cometary-like, and interpreted to be pointing away from the high mass star and thus being the smoking gun of a pulsar wind. These results were found to be consistent with observations by the MAGIC collaboration, simultaneously using MERLIN in the UK, the European VLBI Network (EVN), and the VLBA in the USA (Albert et al. 2008). The comparison between Dhawan et al. (2006) and Albert et al. (2008) images at the same orbital phase (but obtained 10 orbital cycles apart) show a high degree of similarity on both its morphology and flux, which suggests a certain stability of the physical processes involved in the radio emission. 
The tail is not always seemingly pointing in the {\it right} direction: as pointed for instance by Romero et al. 2007 or 
Zdiarszki et al. 2010, a high spin-down power pulsar would generate a wind which should overcome the stellar wind power, 
generating a flat inter-wind shock, or even wrapping it around the Be star, not the pulsar. 
Morphology-wise, the flip-flopping model may alleviate this problem, since even for a high $\dot E$ pulsar, when a propeller is running, there is no pulsar wind and the high-energy emission is not coming from the inter-wind shock. The analysis of the conical shape of the radio emission would not apply. 
MHD simulations of accretion onto a magnetized neutron star in the propeller regime have been recently 
been presented by Romanova et al. (2003) and Toropina et al. (2006).
Their simulations showed that accreting matter is expelled from the equatorial region of the magnetosphere moving away from the star in a supersonic, disk-shaped outflow.  Whether this outflow may be related with the radio morphology is yet an open question.

\subsection{Absence of pulsations at all frequencies}

Finally, we remark that it is natural to expect a failure in detecting pulsations from a magnetar-composed \lsi\ system. Radio pulsations are not common in magnetars. A few examples of radio pulsed emission were detected mostly connected with the outbursts of a few transient objects (Camilo et al. 2006, 2007; Levin et al. 2010), but it is far from being a common or stable property of these highly magnetized neutron stars.

On the other hand, apastron would be the only region of the orbit when, for values of accretion rate so permitting, the pulsar in \lsi\ would have an unscathed magnetosphere. However, the stellar winds might have prevented radio pulsations to be detected because of the strong free-free absorption all over the orbit and/or the large and highly variable dispersion measure (DM) induced by the wind. This would be very similar to the case of PSR B1259-63, which 
at an inter-stellar distance that 
has about the same dimension of the major axis of the orbit of \lsi\  (given the larger 3.4-yr orbit of that system), does not show radio pulsations at periastron either.
In X-rays,  the upper limits derived by Rea et al. (2010) using {\it Chandra} observations at phases close to apastron, mimic the situation of PSR\,B1259-63 (see the discussion in Rea et al. 2011), where again no pulsation is found. Furthermore, X-ray pulsations from magnetars can have pulsed fractions as low as a 4\% (e.g., Mereghetti et al. 2007), well below the $\sim10$\% upper limit derived by Rea et al. (2010).
The absence of X-ray pulsations can be also understood in terms of having an X-ray emission which may be dominated by the shock 
rather than being due to emission from the pulsar magnetosphere.
In periastron, and in general, in all phases where matter enters within the light cylinder, 
the magnetosphere is disrupted and pulsating radio emission is halted. X-ray pulsating emission, on the other hand, is also halted in this case because the accreting matter does not reach the surface of the neutron star.

\section{Predictions \& Outlook}

We can summarize a few testable predictions that are inherent to the scenario we have presented. Some of them are precise, and some represent trends on which further study could shed light on the validity of the assumptions made:

\begin{itemize}

\item If one can track the mass-loss rate evolution of the companion star in  \lsi, the TeV emission will be anti-correlated with it. This would be valid at all timescales, e.g.,  both in an orbit-to-orbit basis, as well as in longer super-orbital timescales. 

\item Assuming the persistence in time of the super-orbital radio modulation, if the cyclical increase of the mass-loss rate of the Be star already produce a flip-flopping \lsi\ system as suggested by Fig. \ref{radio-tev}, it would be natural to expect low TeV fluxes near apastron  for super-orbital phases of $\sim 1\pm 0.2$. 

\item Thus, within this model and under the assumption of the maintenance of the cyclical behavior of the accretion rate as inferred by the (somewhat old) radio data, it is predicted that the system should be already visible by TeV instruments. It should be back in the high state since approximately 2010 May -- June, when the system attained super-orbital phase of 0.2 (there is no coverage of the apastron orbital phases at this epoch reported in Acciari et al. 2010) 
and will disappear or have the TeV emission severely lowered again at super-orbital phase $\sim 0.7-0.8$, or about 2012 October. 

\item If there is ever a giant flare observed from the magnetar hosted in \lsi, it will allow the detection of a magnetar-like pulsating tail, which will give us a handle on the rotational period of the neutron star.

\item It would be natural to expect that when the TeV emission is back to normal (appearing near apastron) the GeV to TeV
connection there 
should be different from that in periastron (this could not be tested yet, since the TeV emission was reduced only 8 months
after {\it Fermi} launch, and an orbital-separated spectrum could not be derived with such dataset). In apastron, the TeV-emitting electrons are present, which may lead to the appearance of a second component above 10 GeV in {\it Fermi} data. This should be less visible in periastron, where the maximal energy of electrons is cut at lower energies by the synchrotron losses.

\item It would be natural to expect that when the TeV emission is significantly reduced or completely disappears along the orbit, including near apastron, the radio morphology should change, given that there is no pulsar wind in place and it is instead the propeller what drives it.

\end{itemize}

We emphasize that a real description of these processes is surely 
far from the simplistic approach presented here.  The transitions between the regimes are in fact not modeled, and
will likely depend on how matter is accreted, on whether a temporary accretion disc is formed, and on many other details.
These will probably be only resolved through numerical simulations. 
This paper should rather be taken as
an exploratory work for the plausibility of the idea when confronted with
the rich phenomenology of the system. Under these caveats, 
we have found that the flip-flop magnetar \lsi\ model 
not only seems to be suggested by a close look at the short burst observed by \swift--BAT and in general, at the X-ray variability, but appears
able to qualitative agree with all other multi-frequency observations of the system and may explain the long term behavior of the TeV variability. 
This qualitative ability of the model to accommodate a wide variety of constraints, even where others seem to fail, encourages further study. 
In particular, studies of accretion onto strongly magnetized systems may prove useful.

\acknowledgements

We acknowledge support from the grants AYA2009-07391 and SGR2009-811, as
well as the Formosa Program TW2010005. NR is supported by a Ramon y Cajal Fellowship. PE acknowledges financial support from the Autonomous Region of Sardinia through a research grant under the program PO Sardegna FSE 2007--2013, L.R. 7/2007 ``Promoting scientific research and innovation technology in Sardinia". This work was also subsidized by the National Natural Science Foundation of China via NSFC-10325313,10521001,10733010,11073021 and 10821061, the CAS key Project KJCX2-YW-T03,
and 973 program 2009CB824800. YPC thanks the Natural Science Foundation of China for support via NSFC-11103020 and 11133002.
We acknowledge W. Bednarek, R. Dubois, G. Dubus, C. Neish, and A. Okazaki  for useful comments.
We also acknowledge the use of public \swift\ and \rxte\ data obtained from the High Energy Astrophysics Science Archive Research Center (HEASARC), provided by NASA's Goddard Space Flight Center


\begin{thebibliography}{99}

\bibitem[]{} Abdo, A., et al. 2009a, ApJ 701, 123
\bibitem[]{} Abdo A. et al., 2009b, ApJ 706, L56
\bibitem[]{} Abdo, A., et al. 2011b, arXiv:1103.4108, 	
submitted to ApJ Letters 
\bibitem[]{} Aharonian F. A., Belyanin A. A., Derishev E. V., Kocharovsky V. V.,
Kocharovsky V. V., 2002, Phys. Rev. D, 66, 023005
\bibitem[]{} Aharonian, F., et al. A\&A, 442, 1
\bibitem[]{} Aharonian, F., et al. 2007, A\&A 469, L1
\bibitem[]{} Albert, J., et al. 2006, Science, 312, 1771
\bibitem[]{} Albert J. et al. 2008, ApJ 684, 1351
\bibitem[]{} Albert, J., et al. 2009, ApJ 693, 303
\bibitem[]{} Acciari V. A. et al. 2008, ApJ 679, 1427
\bibitem[]{} Acciari V. A. et al. 2009, ApJ 700, 1034 
\bibitem[]{} Acciari V. A. et al. 2010, ApJ in press, arXiv:1105.0449
\bibitem[]{} Anderhub, H., et al. 2009, ApJ, 706, L27
\bibitem[]{} Aptekar, R. L., 2001, ApJS 137, 227

\bibitem[]{} Aragona C., McSwain M. V., Grundstrom E. D., Marsh A. N., Roettenbacher R. M., Hessler K. M., Boyajian T. S., \& Ray P. S., 2009, ApJ, 698,
514

\bibitem[]{} Barthelmy S. D. et al. GCN 8215, 2008
\bibitem[]{} Bednarek, W. 2009, MNRAS 397, 1420
\bibitem[]{} Bednarek, W. \& Pabich, J. 2011, MNRAS 411, 1701
\bibitem[]{} Bongiorno, S. D., Falcone, A. D., Stroh, M., Holder, J., Skilton, J. L., Hinton, J. A., 
Gehrels, N., \& Grube, J., 2011, arXiv: 1104.4519, submitted to ApJ Letters
\bibitem[]{} Bozzo, E., Falanga, M., \& Stella, L. 2008, ApJ 683, 1031
\bibitem[]{} Bosch-Ramon A., Paredes J. M., Romero G. E., \& Rib\'o M.  2006, A\&A 459, L25
\bibitem[]{} Bosch-Ramon V., \& Khangulyan D. 2009, International Journal of Modern Physics D18, 347

\bibitem[]{} Camilo, F., et al. 2006, Nature, 442, 892
\bibitem[]{} Camilo, F., Ransom, S. M., Halpern, J. P., Reynolds, J. 2007, ApJ, 666, L93

\bibitem[]{} Campana, S., Stella, L., Mereghetti, S., \& Colpi, M. 1995, A\&A, 297, 385
\bibitem[]{} Casares, J., Ribas, I., Paredes, J. M., Mart\'i, J., \& Allende Prieto, C. 2005,
MNRAS, 360, 1105
\bibitem[]{} Castor, J. I., \& Lamers, H. J. G. L. M. 1979, ApJ, 39, 481
\bibitem[]{} Chernyakova M., Neronov A., Lutovinov A., Rodriguez J., Johnston S., 2006, MNRAS, 372, 1585 
\bibitem[]{} Chernyakova M., Neronov A., Aharonian F., Uchiyama Y., Takahashi T., 2009, MNRAS, 397, 2123
\bibitem[]{} Corbet, R., et al. 2011, ATel 3221

\bibitem[]{} Davies, R. E., \& Pringle, J. E. 1981, MNRAS, 196, 209
\bibitem[]{} Dhawan, V., Mioduszewski, A., \& Rupen, M. 2006, in Proc. of the VI
Microquasar Workshop, Como-2006
\bibitem[]{} de Pasquale M. et al. GCN 8209, 2008
\bibitem[]{} Dickey \& Lockman, 1990, ARAA, 28, 215
\bibitem[]{} Dubus, G. 2006, A\&A 456, 801
\bibitem[]{} Dubus G. 2006b, A\&A 456, 80
\bibitem[]{} Dubus G. \& Giebels B.,  ATel 1715, 2008
\bibitem[]{} Dubus, G. 2010, ASP Conference Series 422, 23; Proceedings of the workshop on High Energy Phenomena in Massive Stars.
\bibitem[]{} Dubus, G., Cerutti, B, \& Henri, G. 2010, A\&A 516, A18
\bibitem[]{} Duncan, R. C., \& Thompson C. 1992, ApJ 392, L9



\bibitem[]{} Esposito, P., Caraveo, P. A., Pellizzoni, A., de Luca, A., Gehrels, N., Marelli, M. A.,  2007, A\&A 474, 575
\bibitem[]{} Evans P. A. et al. GCN 8211, 2008


\bibitem[]{} Falcone, A. D., et al 2010, ApJ 708, L52

\bibitem[]{} Hinton, J., et al. 2009, ApJ 690, L101

\bibitem[]{} Falcke, H., \& Biermann, P. L. 1995, A\&A, 293, 665

\bibitem[]{} Gavriil, F. P., Kaspi, V. M., \& Woods, P. M. 2002, Nature, 419, 142
\bibitem[]{} Gavriil, F. P., et al. 2008, Science 319, 1802
\bibitem[]{} Gehrels, N. et al. 2004, ApJ 611, 1005
\bibitem[]{} Gnusareva, V. S., \& Lipunov, V. M. 1985, Sov. Astron. 29, 645

\bibitem[]{}  Gogus E., Woods, P.M., Kouveliotou, C., van Paradijs, J., Briggs, M.S., Duncan, R.C., \& Thompson, C. 1999, ApJ, 526, L93

\bibitem[]{}  Gogus, E., Kouveliotou, C.,, Woods, P.M., Thompson, C., Duncan, R.C. \& Briggs, M.S. 2001, ApJ, 558, 228


\bibitem[]{} Gregory P.C., Huang-Jian Xu, Bachhouse C.J., Reid A., 1989, ApJ 339, 1054
\bibitem[]{} Gregory P. C. 1999, ApJ 520, 361
\bibitem[]{} Gregory, P. C. 2002, ApJ, 575, 427
\bibitem[]{} Gregory, P.C., \& Neish, C., 2002, ApJ, 580, 1133
\bibitem[]{} Grundstrom, E. D., et al. 2007, ApJ, 656, 437


\bibitem[]{}Hadasch D. for the {\it Fermi}-LAT collaboration 2010, presented at the 3rd {\it Fermi} Symposium, Rome, May 10-14
\bibitem[]{}Hilditch R.W. 2001, 'An Introduction to Close Binary Stars', Cambridge University Press
\bibitem{} Hinton, J., et al. 2009, ApJ 690, L101

\bibitem[]{} Illarionov, A. F., \& Sunyaev, R. A. 1975, A\&A, 39, 185
\bibitem[]{} Israel, G. L., et al. 2008, ApJ 685, 1114
\bibitem[]{} Israel, G. L.  \& Dall'Osso, S.  2011, High-Energy Emission from Pulsars and their Systems, Astrophysics and Space Science Proceedings, Springer-Verlag Berlin Heidelberg, 2011, p. 279

\bibitem[]{}Johnston S., Manchester R. N., Lyne A., Bailes M., Kaspi V. M., Qiao G., DÕAmico N., 1992, ApJ, 387, L37 
\bibitem[]{}Johnston S., Manchester R. N., McConnell D., Campbell-Wilson D., 1999, MNRAS, 302, 277 
\bibitem[]{}Johnston S., Ball L., Wang N., Manchester R. N., 2005, MNRAS, 358,
1069



\bibitem[]{} Khangulyan D., Hnatic, S., Aharonian F. A., \& Bogovalov S. 2007, MNRAS 380, 320
\bibitem[]{} Khangulyan D., Aharonian F. A., \& Bosch-Ramon V. 2008 MNRAS 383, 467
\bibitem[]{} Kumar, H. S., Ibrahim, A. I., Safi-Harb, S., 2010, ApJ 716, 97

\bibitem[]{} Kumar, H. S., Safi-Harb, S., 2008, ApJ 678, L43


\bibitem[]{} Lamers, H. J. G. L. M., \& Cassinelli, J. P. 1999, Introduction to Stellar Winds (Cambridge: Cambridge Univ. Press)
\bibitem[]{} Levin, L., et al. 2010, ApJ 721, L33
\bibitem[]{} Li, J., Torres D. F., Zhang, S., Chen, Y., Hadasch, D., Ray, P. S., Kretschmar, P., Rea, N., \& Wang, J. 2011, ApJ 733, 89 
\bibitem[]{} Lipunov V. M., 1987, Ap\&SS 132, 1 
\bibitem[]{} Lipunov V. M., Nazin, S. N., Osminkin e. Yu., \& Prokhorov M. E. 1994, A\&A 282, 61

\bibitem[]{} Maraschi L. \& Treves A. 1981, MNRAS, 194, 1
\bibitem[]{} Marlborough J. M., Zijlstra J.-W., Waters L. B. F. M., 1997, A\&A, 321,
867
\bibitem[]{} Mart\'i J., Paredes J. M., 1995, A\&A, 298, 151
\bibitem[]{} Mennickent, R. E., Vogt, N., Barrera, L. H., Covarrubias, R., \& Ram\'irez,
A. 1994, A\&AS, 106, 427
\bibitem[]{} Mereghetti, S. 2007, Ap\&SS, 308, 13.
\bibitem[]{} Mereghetti, S. 2008, A\&AR, 15, 225
\bibitem[]{} Mereghetti, S. et al. 2009, ApJ 696, L74
\bibitem[]{} Mirabel, I. F., 2006, Science, 312, 1759,
\bibitem[]{} Mu\~noz-Arjonilla A. J., et al. ATel 1740, 2008
\bibitem[]{} Mu\~noz-Arjonilla, A. J., Mart', J., Combi, J. A., Luque-Escamilla, P., S‡nchez-Sutil, J. R., Zabalza, V., Paredes, J. M. 2009, A\&A, 497, 457



\bibitem[]{} Okazaki, A. T., \& Negueruela, I.,  2001, A\&A 377, 161
\bibitem[]{} Okazaki, A. T., et al., 2002, MNRAS, 337, 967
\bibitem[]{} Okazaki, A. T. 2010, presented at the Workshop on Galactic Variable Sources,
Heidelberg, December 1--4, 2010.
\bibitem[]{} Okazaki, A. T., Nagataki, S., Naito, T., Kawachi, A., Hayasaki, K., Owocki, S. P.; Takata, J., 2011
PASJ in press, arXiv:1105.1481
\bibitem[]{} Orellana, M.,  Romero, G. E.,  Okazaki A. T. \&  Owocki, S. P.
Proceedings of the Asociaci\'on Argentina de Astronom\'ia BAAA, 50, 2007 
G. Dubner, M. Rovira, A. Piatti \& F. A. Bareilles, eds.
\bibitem[]{} Orellana, M., \& Romero, G.E., 2007, Ap\&SS, 309, 333


\bibitem[]{} Paredes J. M., Ribo, M., Bosch-Ramon, V., West, J. R., \& Butt, Y. M. 2007, ApJ 664, L39

\bibitem[]{}  Perna, R., \& Pons, J. 2011, ApJ 727, L51
\bibitem[]{} Ray, P. S., et al. 2008, ATel, 1730
\bibitem[]{} Rea, N., Torres, D. F., ATel 1731, 2008
\bibitem[]{} Rea, N. et al. 2009, MNRAS, 396, 2419
\bibitem[]{} Rea, N., Torres, D. F., van der Klis, M., Jonker, P. G., M\'endez, M., \& Sierpowska-Bartosik, A. 2010, MNRAS 405, 2206
\bibitem[]{} Rea, N., et al. 2010, Science, 330, 944
\bibitem[]{} Rea, N., Torres, D. F., A. G. Caliandro,  Hadasch, D., van der Klis, M., Jonker, P. G., M\'endez, M., \& Sierpowska-Bartosik, A. 2011, MNRAS in press, arXiv:1105.5585
\bibitem[]{} Rea \& Esposito 2011, `High-Energy Emission from Pulsars and their Systems: Proceedings of the First Session of the Sant Cugat Forum on Astrophysics', Springer-Verlag Berlin Heidelberg, 2011, p. 247
\bibitem[]{} Rea, N., Torres, D. F., 2011, ApJ Letters 737, 12

\bibitem[]{} Romanova, M.M., Toropina, O.D., Toropin, Yu.M., \& Lovelace, R.V.E. 2003, ApJ, 588, 400

\bibitem[]{} Romero, G.E. et al., 2007, A\&A, 474, 15 


\bibitem[]{} Scargle 1998, ApJ, 504, 405

\bibitem[]{}Sidoli, L.;, Pellizzoni, A., Vercellone, S., Moroni, M., Mereghetti, S., Tavani, M., 2006, MNRAS 459, 901
\bibitem[]{} Sierpowska-Bartosik A. \& Torres D. F. 2007, ApJ Letters, 671, 145
\bibitem[]{} Sierpowska-Bartosik A. \& Torres D. F. 2008, Astroparticle Physics 30, 239
\bibitem[]{} Sierpowska-Bartosik A. \& Torres D. F. 2009, ApJ 693, 1462
\bibitem{} 	Skilton, J. L. et al. 2009, MNRAS 399, 317



\bibitem[]{} Tam, P. H.,  et al. 2011, arXiv:1103.3129, submitted to ApJ Letters
\bibitem[]{} 
Thompson, C. \& Duncan, R. C. 1993, ApJ 408, 194

\bibitem[]{}  Thompson, C., Lyutikov, M., \& Kulkarni, S. R. 2002, ApJ, 574, 332 


\bibitem[]{} Toropin, Yu., M., Toropina, O. D., \& Lovelace, R. V. E., 2006, Proceedings of YSC-13 YSC Proceedings, Vol. 1, 2006 A. Golovin, A. Simon Eds.
\bibitem[]{} Torres, D. F., 2011, in
`High-Energy Emission from Pulsars and their Systems: Proceedings of the First Session of the Sant Cugat Forum on Astrophysics', Springer-Verlag Berlin Heidelberg, 2011, p. 531,  arXiv:1008.0483
\bibitem[]{} Torres, D. F., for the {\it Fermi}-LAT collaboration 2010, presented at the Workshop on Galactic Variable Sources,
Heidelberg, December 1--4, 2010.
\bibitem[]{} Torres, D. F., Zhang, S., Li, J., Rea, N., Caliandro, G. A., Hadasch, D., Chen, Y., Wang, J. \& Ray P. S., 
2010, ApJ Letters 719, 104
\bibitem[]{} Trushkin, S. \& Nizhelskij, N. 2010, Poster at the 38th COSPAR Scientific Assembly 2010, Accretion on Compact Objects and Fast Phenomena in Multiwavelength Era (E13), 	
Symposium E, session 13, paper number E13-0075-10 (Poster, Nr. Wed-152)

\bibitem[]{} Tueller, M. et al. 2010, ApJS, 186, 378
\bibitem[]{} Waters, L. B. F. M. 1986, A\&A, 162, 121


\bibitem[]{}  Woods, P. M., et al. 2004, ApJ 605, 378
\bibitem[]{}  Woods, P. M. \& Thompson C., 2006, Compact stellar X-ray sources. Edited by Walter Lewin \& Michiel van der Klis. Cambridge Astrophysics Series, No. 39. Cambridge University Press, p. 547 - 586


\bibitem[]{} Zamanov R. K., 1995, MNRAS 272, 308
\bibitem[]{} Zamanov R. K., Mart\'i, J., Paredes J. M. et al. 1999, A\&A 351, 543
\bibitem[]{} Zamanov R. K., \& Mart\'i, J. 2000, A\&A Letters 358, 55
\bibitem[]{} Zamanov R. K., Mart\'i J. M., \& Marziani P. 2001, arXiV 0110114
\bibitem[]{} Zdziarski, A. A., Neronov, A., \& Chernyakova, M. 2010, MNRAS 403, 1873
\bibitem[]{} Zhang, S., Torres, D. F., Li, J., Chen, Y. P., Rea, N., \& Wang, J. M. 2010, MNRAS 408, 642

\end{thebibliography}
\end{document}